\documentstyle[psfig,cite]{lamuphys}
\makeatletter
\let\chapter\hid@chapter
\makeatother
\begin{document}
\pagenumbering{arabic}
\title{Driven Tunneling: Chaos and Decoherence}

\author{Peter H\"anggi\inst{1}, Sigmund Kohler\inst{2}, and
Thomas Dittrich\inst{3}}

\institute{Institut f\"ur Physik, Universit\"at Augsburg,
D-86135 Augsburg, Germany
\and
Depto.\ de F\'\i sica Te\'orica de la Materia Condensada,
Universidad Aut\'onoma, E-28049 Madrid, Spain
\and
Depto.\ de F\'\i sica, Universidad de los Andes, Santaf\'e de Bogot\'a,
Colombia}

\maketitle

\begin{abstract}
Chaotic tunneling in a driven double-well system is investigated in absence as
well as in the presence of dissipation. As the constitutive mechanism of
chaos-assisted tunneling, we focus on the dynamics in the vicinity of
three-level
crossings in the quasienergy spectrum. They are formed when a tunnel doublet,
located on a pair of symmetry-related tori in the classical phase space,
approaches a chaotic singlet in energy. The coherent quantum dynamics near the
crossing, in particular the enhanced tunneling that involves the chaotic
singlet state as a ``step stone'', is described satisfactorily by a
three-state model. It
fails, however, for the corresponding dissipative dynamics, because incoherent
transitions due to the interaction with the environment indirectly couple the
three states in the crossing to the remaining quasienergy states. We model
dissipation by coupling the double well, the driving included, to a heat bath.
The time dependence of the central system, with a quasienergy spectrum
containing exponentially small tunnel splittings, requires special
considerations when applying the Born-Markov and rotating-wave
approximations to derive a master equation for the density operator.
We discuss the effect of decoherence on the now transient chaos-assisted
tunneling: While decoherence
is accelerated practically independent of temperature near the center of
the crossing, it can be stabilzed with increasing temperature at a
chaotic-state induced exact crossing of the ground-state quasienergies.
Moreover the asymptotic amount of coherence left within the vicinity of the
crossing is enhanced if the temperature is below the splitting of the avoided
crossing; but becomes diminished when temperature raises above the splitting
(chaos-induced coherence or incoherence, respectively). The asymptotic state
of the driven dissipative quantum dynamics partially resembles the, possibly
strange, attractor of the corresponding damped driven classical dynamics, but
also exhibits characteristic quantum effects.
\end{abstract}
%

\section{Introduction}
\label{ddw:sec:intro}

The interplay of classical chaos and dissipation in a quantum system bears
interesting effects at the border between classical and quantum mechanics
like, e.g., the suppression of classical chaos by quantum interference
\cite{CasatiChirikovIzrailevFord79}
or its restauration by dissipation \cite{DittrichGraham90}.
While the mutual influence of quantum coherence and
classical chaos is under investigation since many years,
the additional effects caused by coupling the chaotic system to an
environment, namely dissipation and decoherence, have been studied only rarely.
One reason is that by including dissipation, the
computational effort grows drastically, since one has to deal with density
matrices instead of wave functions.

In classical Hamiltonian systems, the transition from regular motion to
chaos is most clearly visible in the change of the phase-space structure:
With increasing nonlinearity, regular tori
successively dissolve in adjacent chaotic layers which grow in size and merge
until the whole phase space is uniformly covered by a chaotic sea where the
dynamics is locally hyperbolic and globally diffusive \cite{Ott93}.
Research in quantum chaos has initially been concentrated on this limiting case
of ``hard chaos'', because the absence of structure in phase
space facilitates the description.

Closer to the generic situation, however, is the intermediate regime with an
extremely intricate interweaving of regular and chaotic areas, as described by
the Kolmogorov-Arnol'd-Moser (KAM) theorem, with self-similar hierarchies of
regular islands. It is in this regime that we expect the most interesting, but
at the same time least tractable, phenomena of chaos-coherence interplay
to occur. A prominent example is chaotic tunneling, the coherent exchange of
probability between symmetry-related regular regions that are separated
dynamically by a chaotic layer, instead of a static potential barrier
\cite{DavisHeller81,BohigasTomsovicUllmo90a,BohigasTomsovicUllmo90b,%
BohigasTomsovicUllmo93,TomsovicUllmo94,LinBallentine90,LinBallentine92,%
UtermannDittrichHanggi94,HanggiUtermannDittrich94,%
LatkaGrigoliniWest94a,LatkaGrigoliniWest94b,LatkaGrigoliniWest94c,%
ZanardiGutierrezGomez95,DoronFrischat95,FrischatDoron98,%
LeyvrazUllmo96,RoncagliaBonciIzrailevWestGrigolini94}.
Chaotic tunneling comes about by the simultaneous action of classical
nonlinear dynamics and quantum coherence. Tunneling is extremely sensitive
to any
disruption of coherence as it occurs due to the unavoidable coupling to the
environment: In presence of dissipation, coherent tunneling becomes a transient
that fades out on the way to an asymptotic state
\cite{CaldeiraLeggett83,GrifoniHanggi98}. This is just one instance of the
general rule that decoherence tends to restore classical behaviour, other
examples being the partial lifting, by dissipation, of the
quantum suppression of chaos, and the appearance of quantum stationary states
that show a close resemblance to corresponding classical strange attractors
\cite{DittrichGraham90}. However, particularly for weak dissipation, more
complicated cross effects occur, such as the strong modification of the
decoherence time by chaotic tunneling.

In this contribution, we investigate the mutual influence of chaotic tunneling
and dissipation for a specific, but nevertheless generic case: a
periodically forced bistable system.
The quartic double well with a harmonic driving will serve as our working
model.
In Section \ref{ddw:sec:model} we introduce its Hamiltonian and the underlying
symmetries. To provide the necessary background, we also briefly review other
important features of this system, in particular driven tunneling and its
coherent suppression and modification in the presence of classical chaos
without damping.

Dealing with a driven system, its quantum dynamics is adequately analyzed in
terms of the Floquet or quasienergy spectrum, also introduced in Section
\ref{ddw:sec:model}. The quasispectrum associated with chaotic tunneling
exhibits a characteristic feature: Quasienergies of chaotic singlets frequently
intersect tunnel doublets which are supported by regular tori. As the basic
mechanism of chaotic tunneling we study, in Sections \ref{ddw:sec:cat} and
\ref{sec:diss}, the coherent and dissipative quantum dynamics in the
vicinity of
such singlet-doublet crossings. While in the coherent case the dynamics is well
described in a three-state approximation, the coupling to the environment
indirectly couples the three states to all other states. On the basis of
numerical results for the full driven double well with dissipation, we reveal
the limitations of the three-level approximation and identify additional
features of the full dynamics not covered by it. In particular, we consider the
long-time asymptotics and the phase-space structure associated with it.

Also on the classical level, the presence of friction has profound consequences
for the phase-space structure: Due to the net contraction of phase-space
volume, stationary states are restricted to manifolds of lower dimensionality
than the underlying phase space. Depending on friction
strength and details of the system, this attractor may be consist of fixed
points, of limit cycles, or, if the classical dynamics is chaotic, of
a strange attractor with self-similar, fractal geometry. On a quantum
level, the structures associated with classical attractors are smeared out on a
scale $\hbar$, yet leave clear traces in the asymptotic state of the
corresponding dissipative quantum dynamics \cite{DittrichGraham87}. We
study the
classical-quantum correspondence of the asymptotic state in
Section~\ref{ddw:sec:attractor}.

\section{The model}
\label{ddw:sec:model}

\begin{figure}[t]
\parbox[t]{6.5cm}{\psfig{width=6.5cm,figure=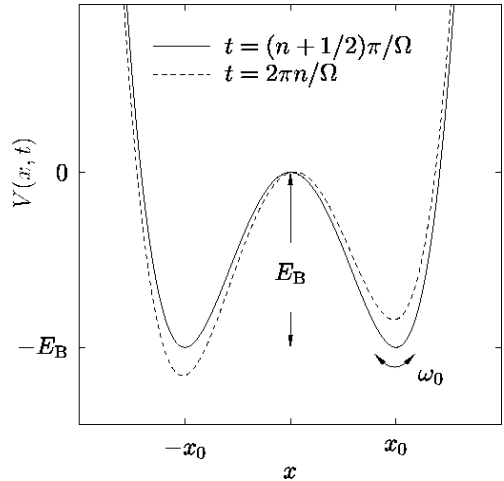}}
\hfill\parbox[b]{5cm}{
\caption{\label{ddw:fig:potential}
Sketch of the driven double well potential described by the time-dependent
Hamiltonian (\protect\ref{ddw:H}) at two different phases.}\vspace{5ex}}
\end{figure}

We consider the quartic double well with a
spatially homogeneous driving force, harmonic in time. It is defined by the
Hamiltonian
\begin{eqnarray}
H(t) &=& H_{\rm DW} + H_F(t) , \label{ddw:H} \\
H_{\rm DW} &=& {p^2\over 2m} - {1\over 4}m\omega_0^2x^2
                + {m^2\omega_0^4\over 64E_{\rm B}}x^4 , \label{ddw:H0}\\
H_F(t) &=& Sx\cos(\Omega t) . \label{ddw:HF}
\end{eqnarray}
The potential term of the static bistable Hamiltonian $H_{\rm DW}$ possesses
two minima at $x=\pm x_0$, $x_0=(8E_{\rm B}/m\omega_0^2)^{1/2}$, separated by
a barrier of height $E_{\rm B}$ (cf.\ Fig.~\ref{ddw:fig:potential}).
The parameter $\omega_0$ denotes the (angular) frequency of small
oscillations near the bottom of each well.
Apart from mere scaling, the classical phase space of $H_{\rm DW}$
only depends on the presence or absence, and the signs, of the $x^2$ and
the $x^4$ term.
Besides that, it has no free parameter.
This is obvious from the scaled form of the classical equations of motion,
\begin{eqnarray}
\dot{\bar x} &=& \bar p , \label{ddw:dotx} \\
\dot{\bar p} &=& {1\over 2}\bar x - {1\over 2}\bar x^3
                 - F\cos(\bar\Omega \bar t), \label{ddw:dotp}
\end{eqnarray}
where the dimensionless quantities $\bar x$, $\bar p$ and $\bar t$ are
given by $x/x_0$, $p/m\omega_0 x_0$ and $\omega_0t$, respectively.
The influence of the driving on the classical phase-space structure is fully
characterized by the rescaled amplitude and frequency of the driving,
\begin{equation}
F = {S\over\sqrt{8m\omega_0^2E_{\rm B}}}\,,
\qquad \bar\Omega={\Omega\over\phantom{_0}\omega_0}\,.
\end{equation}
This implies that the classical dynamics is independent of the barrier
height $E_{\rm B}$.

In the quantum-mechanical case, this is no longer true:
The finite size of Planck's constant results in a finite number
\begin{equation}
D={E_{\rm B}\over\hbar\omega_0}
\end{equation}
of doublets with energy below the barrier top. The classical limit amounts to
$D \to \infty$.
This is evident from the classical scales for position, $x_0$, and
momentum, $m\omega_0 x_0$, introduced above: The corresponding action scale is
$m\omega_0x_0^2$ and therefore, the  position-momentum uncertainty relation in
the scaled phase space $(\bar x,\bar p)$ reads
\begin{equation}
\Delta\bar x\,\Delta\bar p \geq {\hbar_{\rm eff}\over 2},
\end{equation}
where
\begin{equation}
\hbar_{\rm eff}={\hbar\over m\omega_0 x_0^2}={1\over 8D}
\end{equation}
denotes the effective quantum of action.

In the following, we restrict the driving amplitude to moderate values,
such that the difference between the potential minima remains
much smaller than the barrier height. This implies that the bistable character
of the potential is never lost.

\subsection{Symmetries}
\label{ddw:sec:symmetry}

\subsubsection{Time periodicity.}

The Hamiltonian (\ref{ddw:H}) is $P$-periodic, with $P = 2\pi/\Omega$. As a
consequence of this discrete time-translational invariance of $H(x,p;t)$,
the relevant
generator of the quantum dynamics is the Floquet operator
\cite{Shirley65,Sambe73,ManakovOvsiannikovRapoport86,Chu89,QTAD}
\begin{equation}
U = {\cal T}\, {\rm exp} \left( -{{\rm i} \over \hbar}
\int_0^P {\rm d} t \, H(t) \right),
\end{equation}
where ${\cal T}$ denotes time ordering. According to the Floquet theorem, the
adiabatic states of the system are the eigenstates of $U$. They can be written
in the form
\begin{equation}
|\psi_{\alpha}(t)\rangle = {\rm e}^{-{\rm i} \epsilon_{\alpha}t/\hbar}
|\phi_{\alpha}(t)\rangle, \quad
\label{eq:floquetstat}
\end{equation}
with
\begin{eqnarray*}
|\phi_{\alpha}(t + P)\rangle = |\phi_{\alpha}(t)\rangle.
\end{eqnarray*}
Expanded in these Floquet states, the propagator of the driven system reads
\begin{equation}
U(t',t)=\sum_\alpha {\rm e}^{-{\rm i}\epsilon_\alpha (t'-t)/\hbar}
|\phi_\alpha(t')\rangle\langle\phi_\alpha(t)|.
\label{eq:floquetprop}
\end{equation}
The associated eigenphases $\epsilon_{\alpha}$, referred to as quasienergies,
come in classes, $\epsilon_{\alpha,n}=\epsilon_{\alpha}+n\hbar\Omega$, $n = 0,
\pm 1, \pm 2, \ldots$. This is suggested by a Fourier expansion of the
$|\phi_{\alpha}(t)\rangle$,
\begin{eqnarray}
{\displaystyle |\phi_{\alpha}(t)\rangle} &=&
{\displaystyle \sum_n |c_{\alpha,n}\rangle\,
{\rm e}^{-{\rm i} n\Omega t},} \nonumber \\
{\displaystyle |c_{\alpha,n}\rangle} &=&
{\displaystyle {1 \over P} \int_0^P {\rm d} t \,
|\phi_{\alpha}(t)\rangle\,{\rm e}^{{\rm i} n\Omega t}.}
\label{floquet:fourier}
\end{eqnarray}
The index $n$ counts the number of quanta in the driving field. Otherwise, the
members of a class $\alpha$ are physically equivalent. Therefore, the
quasienergy spectrum can be reduced to a single ``Brillouin zone'',
$-\hbar\Omega/2 \leq \epsilon < \hbar\Omega/2$.

Since the quasienergies have the character of phases, they can be ordered only
locally, not globally. A quantity that is defined on the full real axis and
therefore does allow for a complete ordering, is the mean energy
\cite{GrifoniHanggi98,Chu89,QTAD}
\begin{equation}
E_{\alpha}
= {1 \over P} \int_0^P {\rm d} t \,
\langle\psi_{\alpha}(t)|\,H(t)\,|\psi_{\alpha}(t)\rangle
\equiv \langle\langle\phi_{\alpha}(t)|\,H(t)\,|
\phi_{\alpha}(t)\rangle\rangle.
\label{eq:meanen}
\end{equation}
It is related to the corresponding quasienergy by
\begin{equation}
E_{\alpha} = \epsilon_{\alpha} + \langle\langle\phi_{\alpha}(t)
|\,{\rm i}\hbar\frac{\partial}{\partial t}\,|\phi_{\alpha}(t)\rangle\rangle,
\end{equation}
where the outer angle brackets denote the time average over one period of the
driving, as indicated by Eq.~(\ref{eq:meanen}). The second term on the
right-hand side plays the r{\^o}le of a geometric phase accumulated over this
period \cite{GrifoniHanggi98,Moore93}. Without the driving, $E_{\alpha} =
\epsilon_{\alpha}$,
as it should be. By inserting the Fourier expansion (\ref{floquet:fourier}),
the mean energy takes the form
\begin{equation}
E_\alpha = \sum_n(\epsilon_\alpha+n\hbar\Omega)\,
\langle c_{\alpha,n}|c_{\alpha,n}\rangle.
\label{eq:meanen:fourier}
\end{equation}
It shows that the $n$th Floquet channel gives a contribution
$\epsilon_\alpha + n\hbar\Omega$ to the mean energy, weighted by the Fourier
coefficient $\langle c_{\alpha,n}|c_{\alpha,n}\rangle$ \cite{QTAD}.

Quasienergies and Floquet states are obtained numerically by solving the
matrix eigenvalue equation \cite{Shirley65,Chu89,QTAD}
\begin{equation}
\sum_{n'} \sum_{k'} {\cal H}_{n,k;n',k'} c_{n',k'} = \epsilon c_{n,k},
\label{eq:floqueteq}
\end{equation}
equivalent to the time-dependent Schr\"odinger equation. It is derived by
inserting the eigenstates (\ref{eq:floquetstat}) into the Schr\"odinger
equation, Fourier expanding, and using the representation in the eigenbasis
of the unperturbed Hamiltonian, $H_0 |\Psi_k\rangle = E_k
|\Psi_k\rangle$. We introduced the abbreviations
\begin{eqnarray}
{\cal H}_{n,k;n',k'} &=& (E_k - n\hbar\Omega) \delta_{n-n'} \delta_{k-k'}
+\frac{1}{2}\,S\, x_{k,k'} \, (\delta_{n-1-n'} + \delta_{n+1-n'}),
\label{flo:matrix} \\
c_{n,k} &=& \langle\Psi_k|c_n\rangle, \\
x_{k,k'} &=& \langle\Psi_{k}|\,x\,|\Psi_{k'}\rangle.
\label{eq:abbrev}
\end{eqnarray}

\subsubsection{Time-reversal symmetry.}

The energy eigenfunctions of an autonomous Ham\-il\-ton\-ian
with time-reversal symmetry,
\begin{equation}
{\sf T}: \quad x\to x,\quad p\to -p,\quad t\to -t \label{ddw:time_reversal}
\end{equation}
can be chosen as real \cite{Mehta67,Haake91}. Time-reversal invariance
is generally broken by a magnetic field or by an explicit
time-depen\-dence of the Hamiltonian. However, for the sinusoidal shape of the
driving together with the initial phase chosen above, ${\sf T}$ invariance
is retained and the Schr\"odinger operator ${\cal H}(t)=H(t)-{\rm i}\hbar
\partial_t$ obeys ${\cal H}(t)={\cal H}^*(-t)$.
If now $\phi(x,t)$ is a Floquet state in position representation with
quasi\-ener\-gy $\epsilon$, then $\phi^*(x,-t)$ also is a Floquet state
with the same quasienergy.
This means that we can always find a linear combination of these Floquet
states such that $\phi(x,t)=\phi^*(x,-t)$, or in the frequency domain,
$\phi(x,\Omega)=\phi^*(x,\Omega)$,
i.e., the Fourier coefficients of the Floquet states can be chosen real.

\subsubsection{Generalized parity.}

The invariance of $H_{\rm DW}$ under parity ${\sf
P}$: $x\to -x$, $p\to -p$, $t\to t$ is destroyed by any
spatially constant driving force. With the above choice of $H_F(t)$,
however, a more general, dynamical symmetry remains
\cite{GrossmannDittrichJungHanggi91,GrossmannJungDittrichHanggi91,Peres91}.
It is defined by the operation
\begin{equation}
{\sf P}_P:\quad x\to -x,\quad p\to -p,\quad t\to t+P/2
\label{parity}
\end{equation}
and represents a generalized parity acting in the extended phase space spanned
by $x$, $p$, and phase, i.e., time $t\,{\rm mod}\,P$.
While such a discrete symmetry is of minor importance in classical physics,
its influence on the quantum mechanical quasispectrum $\{\epsilon_\alpha(F)\}$
is profound: It devides the Hilbert space in an even and an odd
sector, thus allowing for a classification of the Floquet states as
even or odd. Quasienergies from different symmetry classes may intersect,
while quasienergies with the same symmetry typically form avoided crossings
\cite{Haake91}. The fact that ${\sf P}_P$ acts in the phase space
extended by time $t\,{\rm mod}\,P$, results in a particularity:
If, e.g., $|\phi(t)\rangle$ is an even
Floquet state, then $\exp({\rm i}\Omega t)|\phi(t)\rangle$ is odd, and vive
versa. Thus, two equivalent Floquet states from neighboring Brillouin zones
have opposite generalized parity. This means that a classification of the
corresponding solutions of the Schr\"odinger equation,
$|\psi(t)\rangle=\exp(-{\rm i}\epsilon t/\hbar)|\phi(t) \rangle$, as even
or odd
is meaningful only with respect to a given Brillouin zone.

The invariance of the system under ${\sf P}_P$ is also of
considerable help in the numerical treatment of the Floquet matrix
(\ref{flo:matrix})
\cite{UtermannDittrichHanggi94,HanggiUtermannDittrich94}.
To obtain a complete set of Floquet
states, it is sufficient to compute all eigenvectors of the Floquet
Hamiltonian in the even subspace whose eigenvalues lie in the first two
Brillouin zones.
The even Floquet states are given by the eigenvectors of ${\cal H}_{\rm e}$
from the first Brillouin zone;
the odd Floquet states are obtained by shifting the (even) ones from the
second to the first Brillouin zone, which changes their generalized parity.
Thus, in the even subspace, we have to diagonalize the matrix
\begin{equation}
{\cal H}_{\rm e} \label{ddw:matrix}
=\left(\begin{array}{ccccccc}
\ddots & \vdots  & \vdots & \vdots & \vdots & \vdots & \\
\cdots &  E_{\rm e}+2\hbar\Omega & X_{\rm eo} & 0 & 0 & 0 & \cdots  \\
\cdots & X_{\rm eo} & E_{\rm o}+\hbar\Omega & X_{\rm oe} & 0 & 0 & \cdots  \\
\cdots & 0 & X_{\rm oe} & \quad E_{\rm e}\quad & X_{\rm eo} & 0 & \cdots  \\
\cdots & 0 & 0 & X_{\rm eo} & E_{\rm o}-\hbar\Omega & X_{\rm oe} & \cdots \\
\cdots & 0 & 0 & 0 & X_{\rm oe} & E_{\rm e}-2\hbar\Omega & \cdots  \\
 & \vdots & \vdots & \vdots & \vdots & \vdots & \ddots
\end{array}\right).
\end{equation}
For the same number of Floquet channels, it has only half the dimension
of the original Floquet matrix (\ref{flo:matrix}). The entries in ${\cal
H}_{\rm
e}$ are themselves blocks of infinite dimension, in principle. They read
explicitly
\begin{eqnarray}
E_{\rm e}=
\left(\begin{array}{cccc}
E_0 & 0 & 0 & \cdots \\
0 & E_2 & 0 & \cdots \\
0 & 0 & E_4 & \cdots \\
\vdots & \vdots & \vdots & \ddots
\end{array}\right)
&,\quad &
E_{\rm o}=
\left(\begin{array}{cccc}
E_1 & 0 & 0 & \cdots \\
0 & E_3 & 0 & \cdots \\
0 & 0 & E_5 & \cdots \\
\vdots & \vdots & \vdots & \ddots
\end{array}\right), \\
\nonumber\phantom{A}\\
X_{\rm eo}={S\over 2}
\left(\begin{array}{cccc}
x_{0,1} & x_{0,3} & x_{0,5} & \cdots \\
x_{2,1} & x_{2,3} & x_{2,5} & \cdots \\
x_{4,1} & x_{4,3} & x_{4,5} & \cdots \\
\vdots & \vdots & \vdots & \ddots
\end{array}\right)
&,\quad &X_{\rm oe}={S\over 2}
\left(\begin{array}{cccc}
x_{1,0} & x_{1,2} & x_{1,4} & \cdots \\
x_{3,0} & x_{3,2} & x_{3,4} & \cdots \\
x_{5,0} & x_{5,2} & x_{5,4} & \cdots \\
\vdots & \vdots & \vdots & \ddots
\end{array}\right).\qquad
\end{eqnarray}
Here, $E_{\rm e}$, $E_{\rm o}$, represent the undriven Hamiltonian, and $X_{\rm
eo}$, $X_{\rm oe}$ the driving field $H_1=Sx/2$, decomposed in the even and odd
eigenstates $|\Psi_k\rangle$ of $H_{\rm DW}$, with $E_k$ denoting an
eigenvalue of $H_{\rm DW}$, and $x_{k,k'}$ a matrix element of the position
operator, see Eq.~(\ref{eq:abbrev}).

\subsection{Tunneling, driving, and dissipation}

With the driving $H_F(t)$ switched off, the classical phase space generated
by $H_{\rm DW}$ exhibits the constituent features of a bistable Hamiltonian
system. A separatrix at $E = 0$ forms the border between two sets
of trajectories: One set, with $E < 0$, comes in symmetry-related pairs, each
partner of which oscillates in either one of the two potential minima. The
other set consists of unpaired, spatially symmetric trajectories, with $E > 0$,
which encircle both wells.

Torus quantization of the integrable undriven double well, Eq.~(\ref{ddw:H0}),
implies a simple qualitative picture of its eigenstates: The unpaired tori
correspond to singlets with positive energy, whereas the symmetry-related pairs
below the top of the barrier correspond to degenerate pairs of eigenstates.
Due to the almost harmonic shape of the potential near its minima, neighboring
pairs are separated in energy approximately by $\hbar\omega_0$. Exact
quantization, however, predicts that the partners of these pairs have small but
finite overlap. Therefore, the true eigenstates come in doublets, each of which
consists of an even and an odd state, $|\Phi_n^+\rangle$ and
$|\Phi_n^-\rangle$, respectively. The energies of the $n$th doublet are
separated
by a finite tunnel splitting $\Delta_n$. We can always choose the global
relative phase such that the superpositions
\begin{equation}
|\Phi_n^{\rm R,L}\rangle = {1\over\sqrt{2}}\left(
|\Phi_n^+\rangle \pm |\Phi_n^-\rangle \right)
\end{equation}
are localized in the right and the left well, respectively.
As time evolves, the states $|\Phi_n^+\rangle$, $|\Phi_n^-\rangle$ acquire
a relative phase $\exp(-{\rm i}\Delta_n t/\hbar)$ and $|\Phi_n^{\rm R}\rangle$,
$|\Phi_n^{\rm L}\rangle$ are transformed into one another after a time
$\pi\hbar/\Delta_n$.
Thus, the particle tunnels forth and back between the wells with a frequency
$\Delta_n/\hbar$. This introduces an additional, purely quantum-mechanical
frequency scale, the tunnel rate
$\Delta_0/\hbar$ of a particle residing in the ground-state doublet.
Typically, tunnel rates are extremely small compared to the frequencies of
the classical dynamics, all the more in the semiclassical regime we are
interested in.

\begin{figure}[t]
\centerline{
\psfig{width=11cm,figure=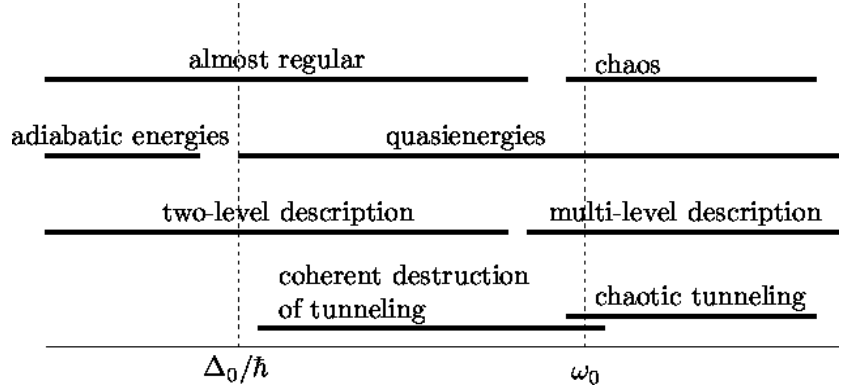}
}
\caption{\label{ddw:fig:timescale}
Tunneling phenomena and the according appropriate levels of description
for the non-dissipative driven double-well potential, Eq.\ (\ref{ddw:H}).
The bars depict the corresponding regimes of the driving frequency $\Omega$.
See Section \ref{ddw:sec:model} for a detailed discussion.
}
\end{figure}

A driving of the form (\ref{ddw:HF}), even if its influence on the classical
phase space is minor, can entail significant consequences for tunneling: It may
enlarge the tunnel rate by orders of magnitude or even suppress tunneling
altogether. For adiabatically slow driving, $\Omega\ll\Delta_0/\hbar$,
tunneling
is governed by the instantaneous tunnel splitting, which is
always larger than its unperturbed value $\Delta_0$ and results in an
enhancement of the tunneling rate \cite{GrossmannJungDittrichHanggi91}.
If the driving is faster, $\Delta_0/\hbar\la\Omega\ll\omega_0$,
cf. Fig.~\ref{ddw:fig:timescale}, the opposite
holds true: The relevant time scale is now given by the inverse of the
quasienergy splitting of the ground-state doublet
$\hbar/|\epsilon_1-\epsilon_0|$.
It has been found \cite{GrossmannJungDittrichHanggi91,GrossmannHanggi92} that
in this case, for finite driving amplitude, $|\epsilon_1-\epsilon_0|<\Delta_0$.
Thus tunneling is always decelerated.
Where the quasienergies of the ground-state
doublet (which are of different generalized parity) intersect as a function
of $F$, the splitting vanishes and tunneling is
brought to a complete standstill by the purely coherent influence of the
driving \cite{GrossmannDittrichJungHanggi91}.

The small energy scales associated with tunneling make it extremely sensitive
to any loss of coherence. As a consequence, the symmetry
underlying the formation of tunnel doublets is generally broken, and an
additional energy scale is introduced, the effective finite width attained by
each discrete level. Tunneling and related coherence phenomena thus
fade out on a time scale $t_{\rm decoh}$.
In general, this time scale gets shorter for higher temperatures,
reflecting the growth of the transition rates (\ref{MasterEquationRWA})
\cite{HanggiTalknerBorkovec90}. However, there exist counterintuitive effects:
For driven tunneling in the vicinity of an exact
crossing of the ground-state doublet, the coherent suppression of
tunneling \cite{GrossmannDittrichJungHanggi91,GrossmannJungDittrichHanggi91,%
GrifoniHanggi98} can be stabilized with higher temperatures
\cite{DittrichOelschlagelHanggi93,OelschlagelDittrichHanggi93,%
DittrichHanggiOelschlagelUtermann95}
until levels outside the doublet start to play a r\^ole.

So far, we have considered only driving frequencies much smaller than the
frequency scale $\omega_0$ of the relevant classical resonances, i.e., a
parameter regime where classical motion is predominantly regular.
In this regime, coherent tunneling is well described within a
two-state approximation \cite{GrossmannHanggi92,GrossmannJungDittrichHanggi91}.
In the dissipative case, however, a two-state approximation must fail
for temperatures $k_{\rm B}T\ga\hbar\omega_0$, where thermal activation to
higher doublets becomes relevant.

\subsection{The onset of chaos}

Driving with a frequency $\Omega\approx\omega_0$ affects also the dynamics of
the classical bistable system, as small oscillations near the bottom of the
wells
become resonant with the driving and classical chaos comes into play
(cf. Fig.~\ref{ddw:fig:timescale}).
In a quantum description, this amounts to resonant multiple excitation of
inter-doublet transitions until levels near the top of the barrier are
significantly populated.

In this frequency regime, switching on the driving has two principal
consequences for the classical dynamics: The separatrix is destroyed as a
closed
curve and replaced by a homoclinic tangle \cite{LichtenbergLiebermann} of
stable
and unstable manifolds. A chaotic layer forms in the vicinity and with the
topology of the former separatrix (cf.\ Fig.~\ref{ddw:fig:poincare}, below).
This opens the way for diffusive transport between the two potential wells.
Due to the nonlinearity of the potential, there is an infinite set of
resonances of the driving with the unperturbed motion, both inside as well as
outside the wells \cite{Escande85,ReichlZheng87}.
Since the period of the unperturbed, closed trajectories diverges for $E \to
0$, both from below and above, the resonances accumulate towards the separatrix
of the unperturbed system. By its large phase-space area, the first resonance
(the one for which the periods of the driving and of the unperturbed
oscillation
are in a 1-to-1 ratio) is prominent among the others and soon (in terms of
increasing amplitude $F$) exceeds in size the ``order-zero'' regular areas
near the bottom of each well \cite{UtermannDittrichHanggi94}.

Both major tendencies in the evolution of the classical phase
space, extension of the chaotic layer and growth of the first
resonance, leave their specific traces in the quasienergy spectrum. The
tunnel doublets characterizing the unperturbed spectrum for $E < 0$ pertain to
states located on pairs of symmetry-related quantizing tori in the regular
regions within the wells. With increasing size of the chaotic layer, the
quantizing tori one by one resolve in the chaotic sea. The corresponding
doublets disappear as distinct structures in the spectrum as they attain a
splitting of the same order as the mean level separation. The gradual widening
of the doublets proceeds as a smooth function of the driving amplitude
\cite{UtermannDittrichHanggi94,HanggiUtermannDittrich94},
which roughly obeys a power law
\cite{Wilkinson86,Wilkinson87,KohlerUtermannHanggiDittrich98}.
As soon as a pair of states is no longer supported by
any torus-like manifold, including fractal \cite{Reichl92} and vague tori
\cite{ShirtsReinhardt82}, the corresponding eigenvalues detach
themselves from the regular ladder to which they formerly belonged.
They can then fluctuate freely in the spectrum and thereby ``collide''
with other chaotic singlets or regular doublets.

The appearance of a regular region, large enough to accommodate several
eigenstates, around the first resonance introduces a second ladder of doublets
into the spectrum. Size and shape of the first resonance vary in a way
different from the main regular region. The corresponding doublet
ladder therefore moves in the spectrum independently of the doublets that
pertain to the main regular region, and of the chaotic singlets. This gives
rise to additional singlet-doublet and even to doublet-doublet encounters.

\section{Chaotic tunneling near singlet-doublet crossings}
\label{ddw:sec:cat}

Near a crossing, level separations deviate vastly, in both directions,
from the typical tunnel splitting (cf.\ Fig.~\ref{ddw:fig:splitting}, below).
This is reflected in time-domain phenomena ranging from the suppression of
tunneling to a strong increase in its rate and to complicated quantum beats
\cite{LatkaGrigoliniWest94a,LatkaGrigoliniWest94b,LatkaGrigoliniWest94c}.
Singlet-doublet crossings, in turn, drastically change the
non-dissipative quasienergy scales and replace the two-level by a three-level
structure. As a consequence, the familiar way tunneling fades out in the
presence of dissipation is also significantly altered. Near a crossing, the
coherent dynamics can last much longer than for the unperturbed doublet,
despite the presence of the same dissipation as outside the crossing,
establishing ``chaos-induced coherence.'' Depending on temperature, it can
also be destroyed on a much shorter time scale.

For the parameters chosen in our numerical studies, higher resonances are
negligible in size. Accordingly, the ``coastal strip'' between the chaotic
layer
along the former separatrix and the regular regions within and outside the
wells, formed by hierarchies of regular islands around higher resonances,
remains narrow (cf.\ Fig.~\ref{ddw:fig:poincare}). For the
tunneling dynamics, the r\^ole of states located in the border region
\cite{DoronFrischat95,FrischatDoron98} is therefore not significant in our
case.
\begin{figure}[t]
\parbox[t]{6cm}{\psfig{width=6cm,figure=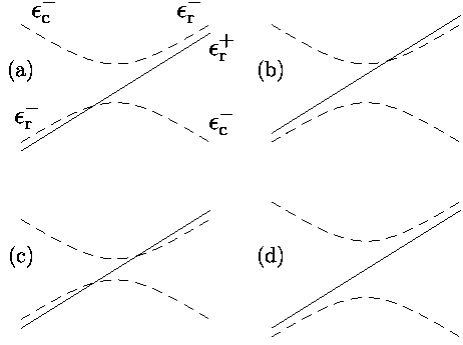}}
\hfill\parbox[b]{5.3cm}{
\caption{ \label{ddw:fig:crossing}
Possible configurations of qua\-sienergy crossings between a chaotic singlet
and a regular doublet. Different line types signify different parity. See
Section~\protect\ref{sec:3s} for the labeling of the levels. Note that only for
configurations (a),(b), the order of the regular doublet is restored in passing
through the crossing. In configurations (c),(d), it is reversed.
}}
\end{figure}

\subsection{Three-level crossings}
\label{sec:3s}
Among the various types of quasienergy crossings that occur according to the
above scenario, those involving a regular doublet and a chaotic singlet are
the most common. In order to give a quantitative account of such crossings and
the associated coherent dynamics, and for later reference in the context of
the incoherent dynamics, we shall now discuss them in terms of a simple
three-state model, devised much in the spirit of
Ref.~\cite{BohigasTomsovicUllmo93}.

Far to the left of the crossing, we expect the following situation: There is a
doublet of Floquet states
\begin{eqnarray}
|\psi_{\rm r}^+(t)\rangle
&=& {\rm e}^{-{\rm i}\epsilon_{\rm r}^+ t/\hbar}|\phi_{\rm r}^+(t)\rangle , \\
|\psi_{\rm r}^-(t)\rangle
&=& {\rm e}^{-{\rm i}(\epsilon_{\rm r}^+ +\Delta) t/\hbar}
|\phi_{\rm r}^-(t)\rangle ,
\end{eqnarray}
with even (superscript $+$) and odd ($-$) generalized parity,
respectively, residing on a pair of quantizing tori in one of the regular
(subscript r) regions. We have assumed the quasienergy splitting $\Delta =
\epsilon_{\rm r}^- - \epsilon_{\rm r}^+$ (as opposed to the unperturbed
splitting) to be positive. The global relative phase is chosen such that
the superpositions
\begin{equation}
\label{eq:rightleft}
|\phi_{\rm R,L}(t)\rangle = \frac{1}{\sqrt{2}}\left(|\phi_{\rm r}^+(t)\rangle
\pm |\phi_{\rm r}^-(t)\rangle\right)
\end{equation}
are localized in the right and the left well, respectively, and
tunnel back and forth with a frequency $\Delta/\hbar$.

As the third player, we introduce a Floquet state
\begin{equation}
|\psi_{\rm c}^-(t)\rangle
= {\rm e}^{-{\rm i}(\epsilon_{\rm r}^+ + \Delta + \Delta_{\rm c})t/\hbar}
|\phi_{\rm c}^-(t)\rangle,
\end{equation}
located mainly in the chaotic (subscript c) layer, so that its time-periodic
part $|\phi_{\rm c}^-(t)\rangle$ contains a large number of harmonics.
Without loss of generality, its parity is fixed to be odd. For the
quasienergy, we assume that $\epsilon_{\rm c}^- = \epsilon_{\rm r}^+ +
\Delta + \Delta_{\rm c} = \epsilon_{\rm r}^- + \Delta_{\rm c}$, where
$|\Delta_{\rm c}|$ can be regarded as a
measure of the distance from the crossing.

The structure of the classical phase space then implies that the mean energy
of the chaotic state should be close to the top of the barrier and far above
that of the doublet. We assume, like for the quasienergies, a small splitting
of the mean energies pertaining to the regular doublet,
$|E_{\rm r}^- - E_{\rm r}^+| \ll E_{\rm c}^- - E_{\rm r}^\pm$.

In order to model an avoided crossing between $|\phi_{\rm r}^-\rangle$ and
$|\phi_{\rm c}^-\rangle$, we suppose that there is a non-vanishing fixed matrix
element
\begin{equation}
b = \langle\langle\phi_{\rm r}^-|H_{\rm DW}|\phi_{\rm c}^-\rangle\rangle > 0.
\end{equation}
For the singlet-doublet crossings under study, we typically find
that $\Delta \ll b \ll \hbar\Omega$. Neg\-lecting the coupling with all
other states, we model the system by the three-state (subscript 3s) Floquet
Hamiltonian
\begin{equation}
{\cal H}_{\rm 3s} = \epsilon_{\rm r}^+
+\left(\begin{array}{ccc}
        0 & 0      &                     0 \\
        0 & \Delta &                     b \\
        0 & b      & \Delta+\Delta_{\rm c}
\end{array}\right)
\label{eq:ham3s}
\end{equation}
in the three-dimensional Hilbert space spanned by $\{|\phi_{\rm
r}^+(t)\rangle, |\phi_{\rm r}^-(t)\rangle, |\phi_{\rm c}^-(t)\rangle\}$. Its
Floquet states are
\begin{eqnarray}
\label{eq:psi012}
|\phi_0^+(t)\rangle
&=& 
|\phi_{\rm r}^+(t)\rangle , \nonumber\\
|\phi_1^-(t)\rangle
&=& 
    \left( |\phi_{\rm r}^-(t)\rangle\cos\beta
    - |\phi_{\rm c}^-(t)\rangle\sin\beta \right) , \\
|\phi_2^-(t)\rangle
&=& 
    \left( |\phi_{\rm r}^-(t)
    \rangle\sin\beta + |\phi_{\rm c}^-(t)\rangle\cos\beta \right) .\nonumber
\end{eqnarray}
with quasienergies
\begin{equation}
\epsilon_0^+ = \epsilon_{\rm r}^+,\quad
\epsilon_{1,2}^- = \epsilon_{\rm r}^+ + \Delta +
\frac{1}{2}\Delta_{\rm c}\mp\frac{1}{2} \sqrt{\Delta_{\rm c}^2+4b^2},
\end{equation}
and mean energies, neglecting contributions of the matrix element $b$,
\begin{eqnarray}
\label{eq:meanen3s}
E_0^+ &=& E_{\rm r}^+ , \nonumber\\
E_1^- &=& E_{\rm r}^-\cos^2\beta + E_{\rm c}^-\sin^2\beta , \\
E_2^- &=& E_{\rm r}^-\sin^2\beta + E_{\rm c}^-\cos^2\beta . \nonumber
\end{eqnarray}

\begin{figure}[p]
\centerline{
\psfig{width=6cm,figure=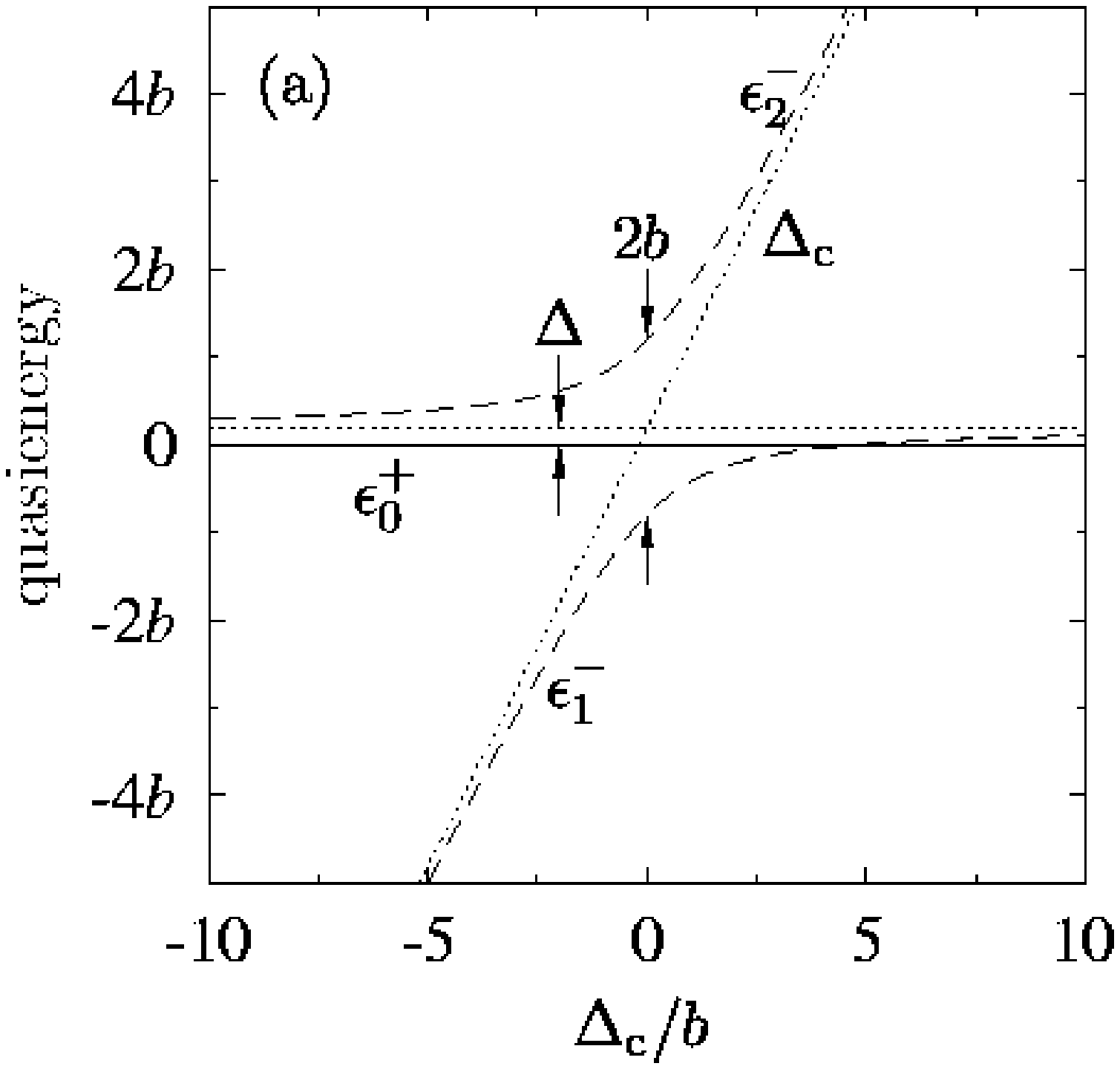}
\psfig{width=6cm,figure=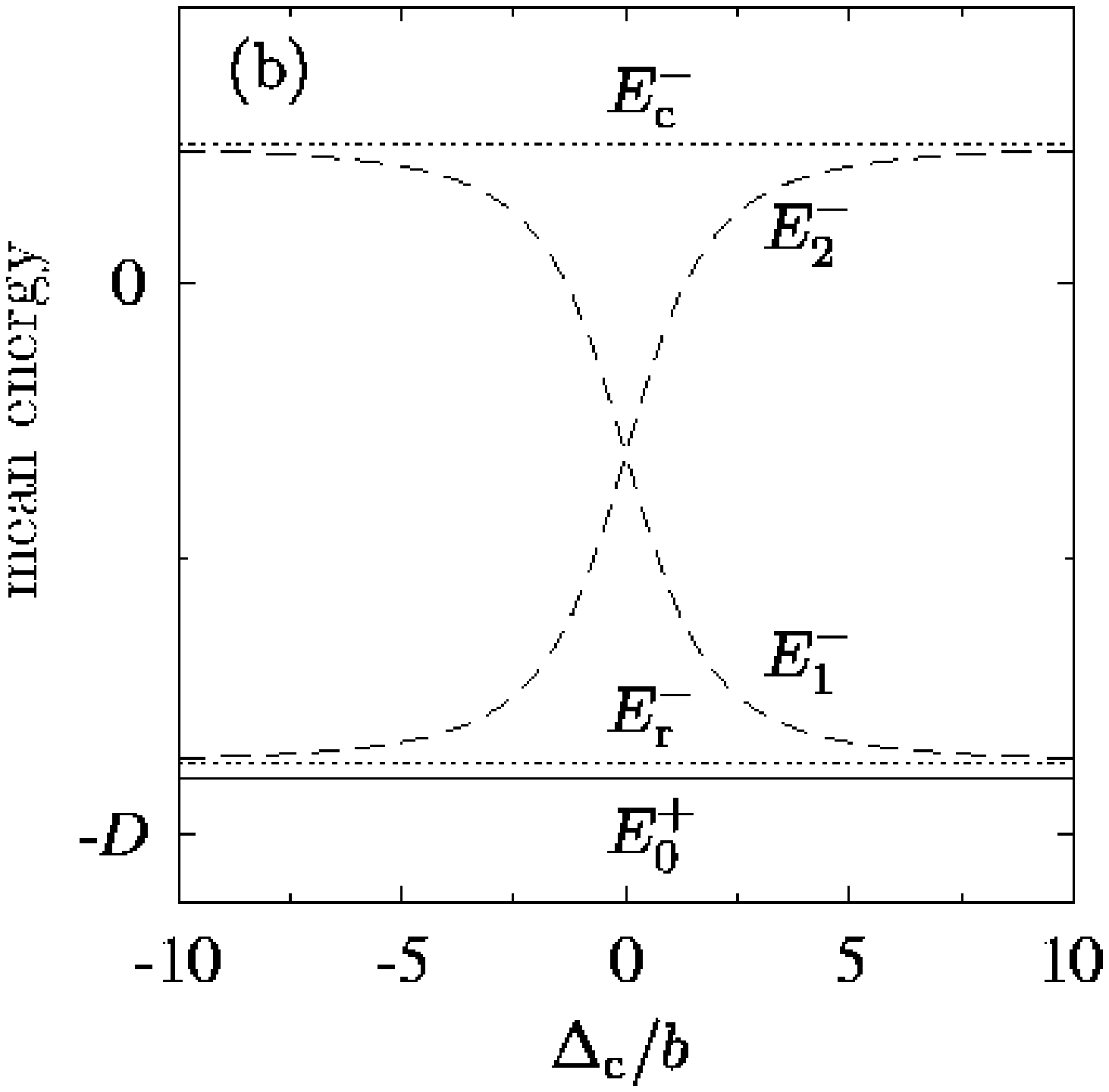}
}
\caption{
A singlet-doublet crossing, according to the three-state model
(\protect\ref{eq:ham3s}), in terms of the quasienergies
(a) and the mean energies (b) as functions of the coupling parameter
$\Delta_{\rm c}/b$. Energies for a corresponding exact crossing (i.e., with the
crossing states uncoupled) are marked by dotted lines, the energies in the
presence of coupling by full and dashed lines for even and odd states,
respectively.
\label{ddw:fig:theo3s} }
\centerline{
\psfig{width=6cm,figure=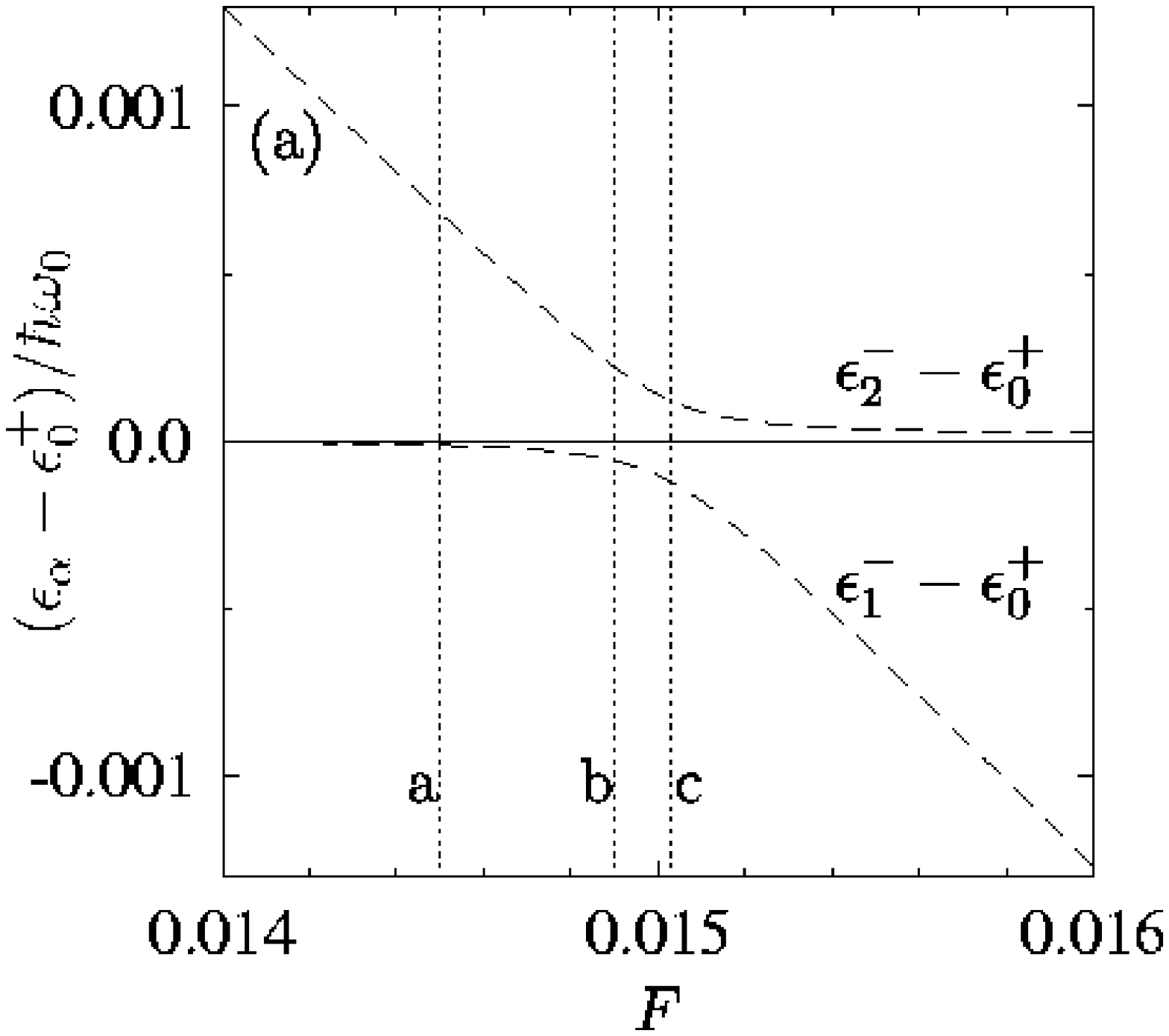}
\psfig{width=6cm,figure=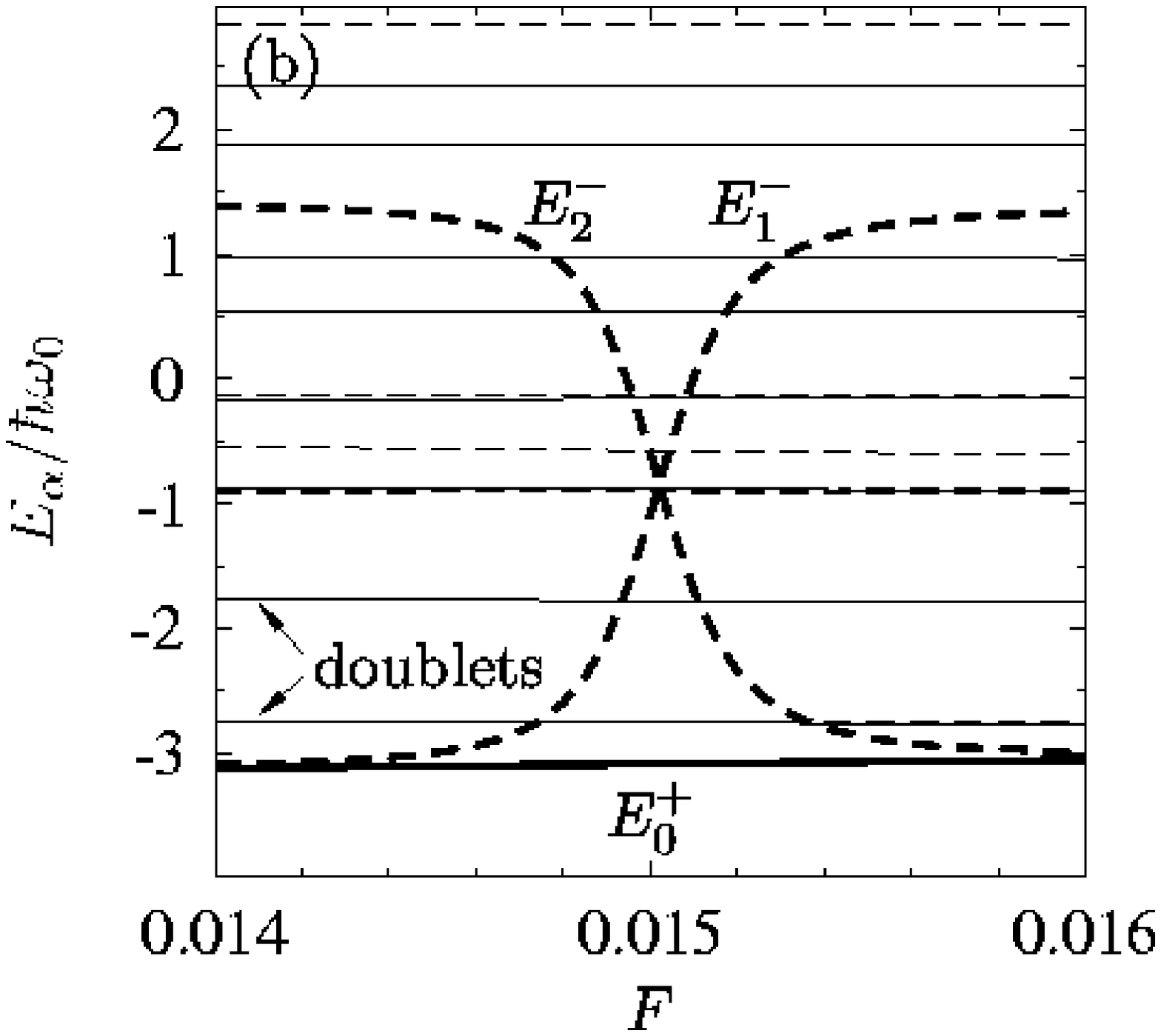}
}
\caption{
Singlet-doublet crossing found numerically for the driven double well,
Eq.\ (\protect\ref{ddw:H}), at $D=4$ and $\Omega = 0.982\,\omega_0$, in
terms of the dependence of the quasienergies (a) and the mean energies (b) on
the driving amplitude $F$. Values of the driving
amplitude used in Fig.\ \protect\ref{ddw:fig:tun3s}
are marked by dotted vertical lines. Full and dashed lines indicate energies
of even and odd states, respectively. Bold lines give the mean energies of
the chaotic singlet and the ground-state doublet depicted in panel (a).
\label{ddw:fig:num3s}
}
\end{figure}%
The angle $\beta$ describes the mixing between the Floquet states $|\phi_{\rm
r}^-\rangle$ and $|\phi_{\rm c}^-\rangle$ and is an alternative measure of the
distance to the avoided crossing. By diagonalizing the Hamiltonian
(\ref{eq:ham3s}), we obtain
\begin{equation}
2\beta = \arctan\left(\frac{2b}{\Delta_{\rm c}}\right), \quad
0 < \beta < \frac{\pi}{2}.
\end{equation}
For $\beta \to \pi/2$, corresponding to $-\Delta_{\rm c} \gg b$, we
retain the situation far left of the crossing, as outlined above, with
$|\phi_1^- \rangle \approx |\phi_{\rm c}^-\rangle$, $|\phi_2^-\rangle \approx
|\phi_{\rm r}^-\rangle$. To the far right of the crossing, i.e., for $\beta
\to 0$ or $\Delta_{\rm c} \gg b$, the exact eigenstates
$|\phi_1^-\rangle$ and $|\phi_2^-\rangle$ have interchanged their phase-space
structure
\cite{LatkaGrigoliniWest94a,LatkaGrigoliniWest94b,LatkaGrigoliniWest94c}. Here,
we have $|\phi_1^-\rangle \approx |\phi_{\rm r}^-\rangle$ and $|\phi_2^-\rangle
\approx |\phi_{\rm c}^-\rangle$.  The mean energy is essentially determined by
this phase-space structure, so that there is also an exchange of $E_1^-$ and
$E_2^-$ in an exact crossing, cf.\ Eq.~(\ref{eq:meanen3s}), while $E_0^+$
remains unaffected (Fig.~\ref{ddw:fig:theo3s}b).
The quasienergies $\epsilon_0^+$ and $\epsilon_1^-$ must intersect close to
the avoided crossing of $\epsilon_1^-$ and $\epsilon_2^-$
(Fig.~\ref{ddw:fig:theo3s}a). Their crossing is
exact, since they pertain to states with opposite parity (cf.\
Fig.~\ref{ddw:fig:crossing}a,b).
\begin{figure}[t]
\centerline{
\psfig{width=7cm,figure=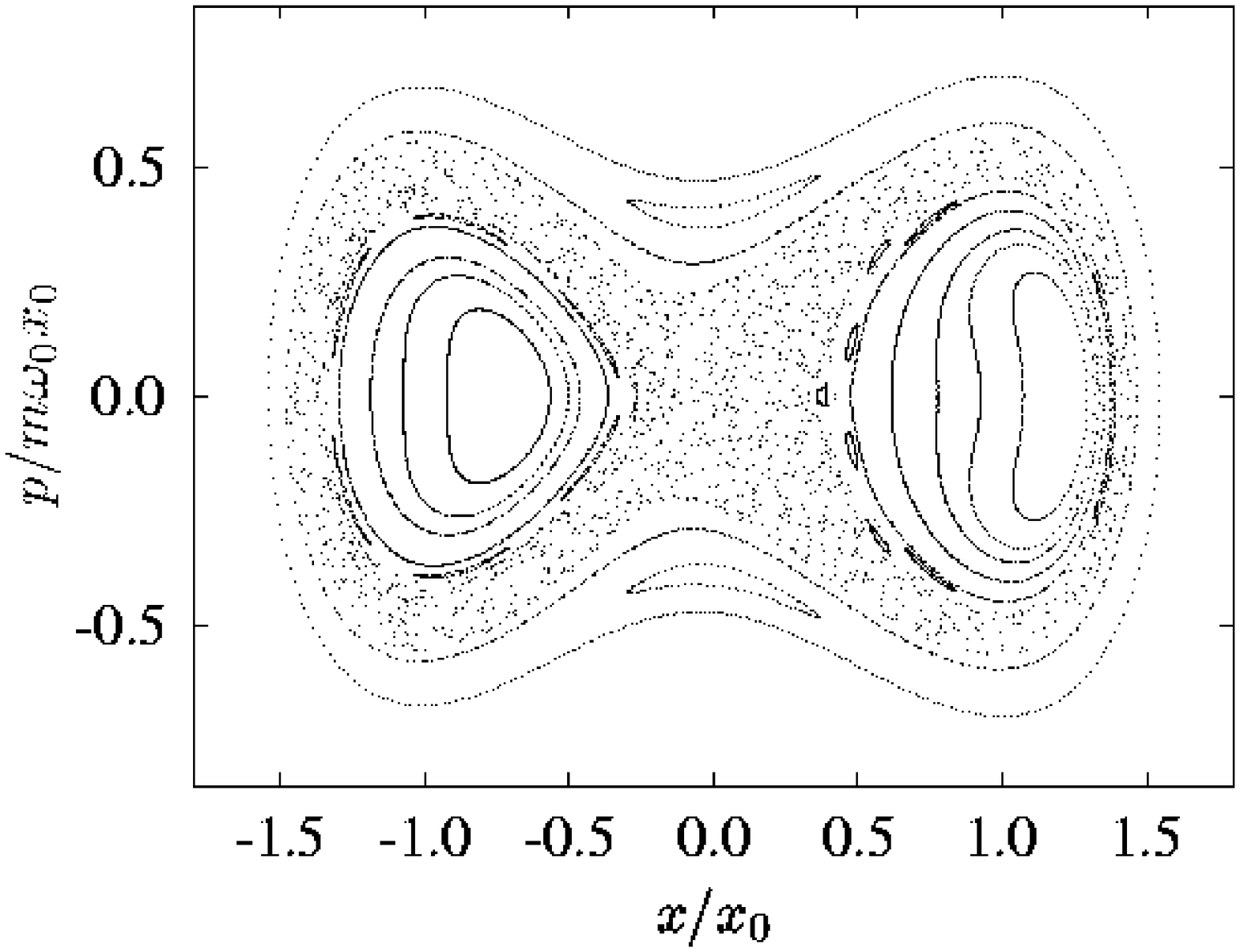}
}
\caption{ \label{ddw:fig:poincare}
Classical stroboscopic phase space portrait, at $t=2\pi n/\Omega$, of the
harmonical\-ly driven quartic double well, Eq.\ (\protect\ref{ddw:H}).
The driving para\-me\-ters $F=0.015$, $\Omega=0.982\,\omega_0$, are at the
center of the singlet-doublet crossing shown in Fig.\
\protect\ref{ddw:fig:num3s}.}\vspace{2ex}
\centerline{
\psfig{width=6cm,figure=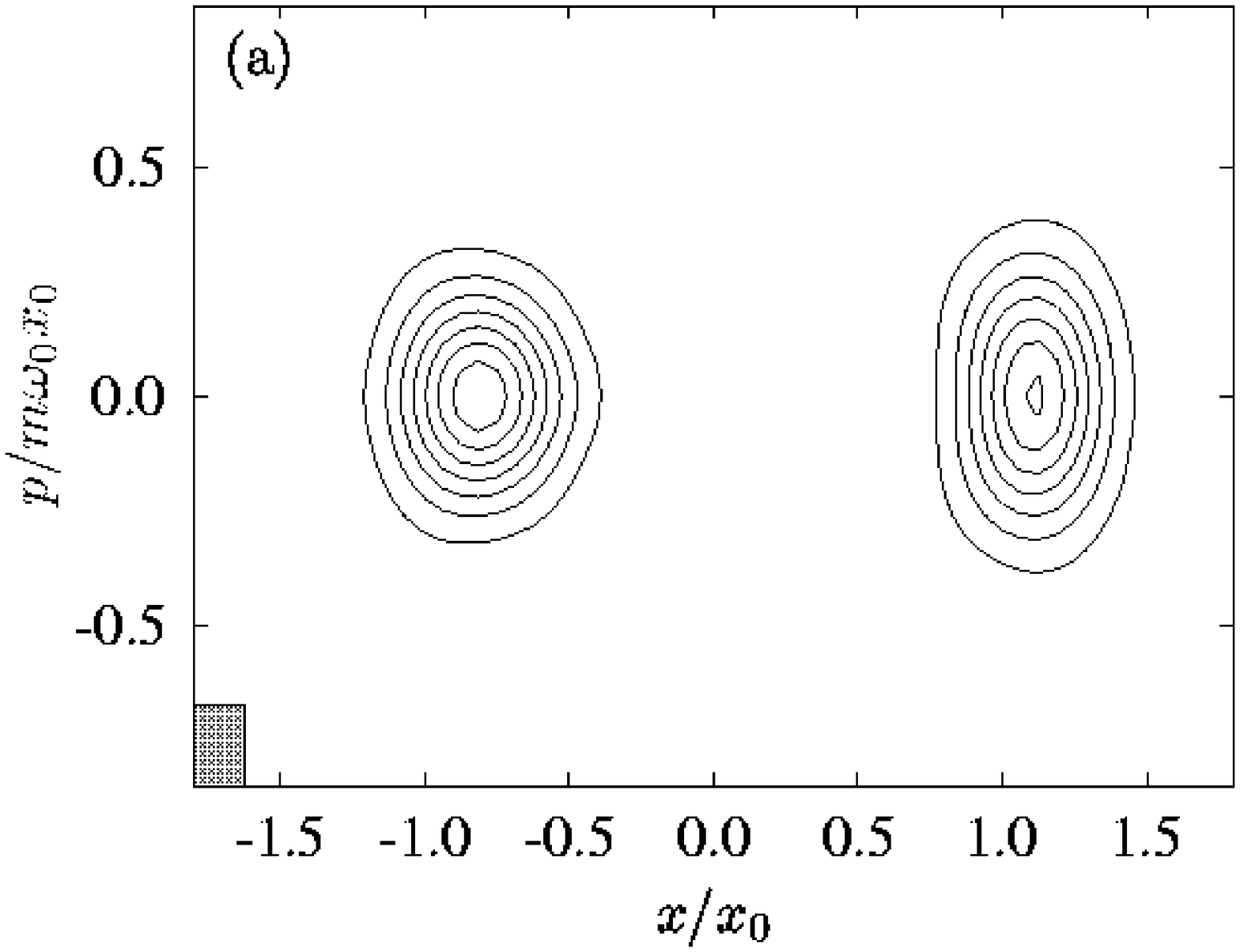}
\psfig{width=6cm,figure=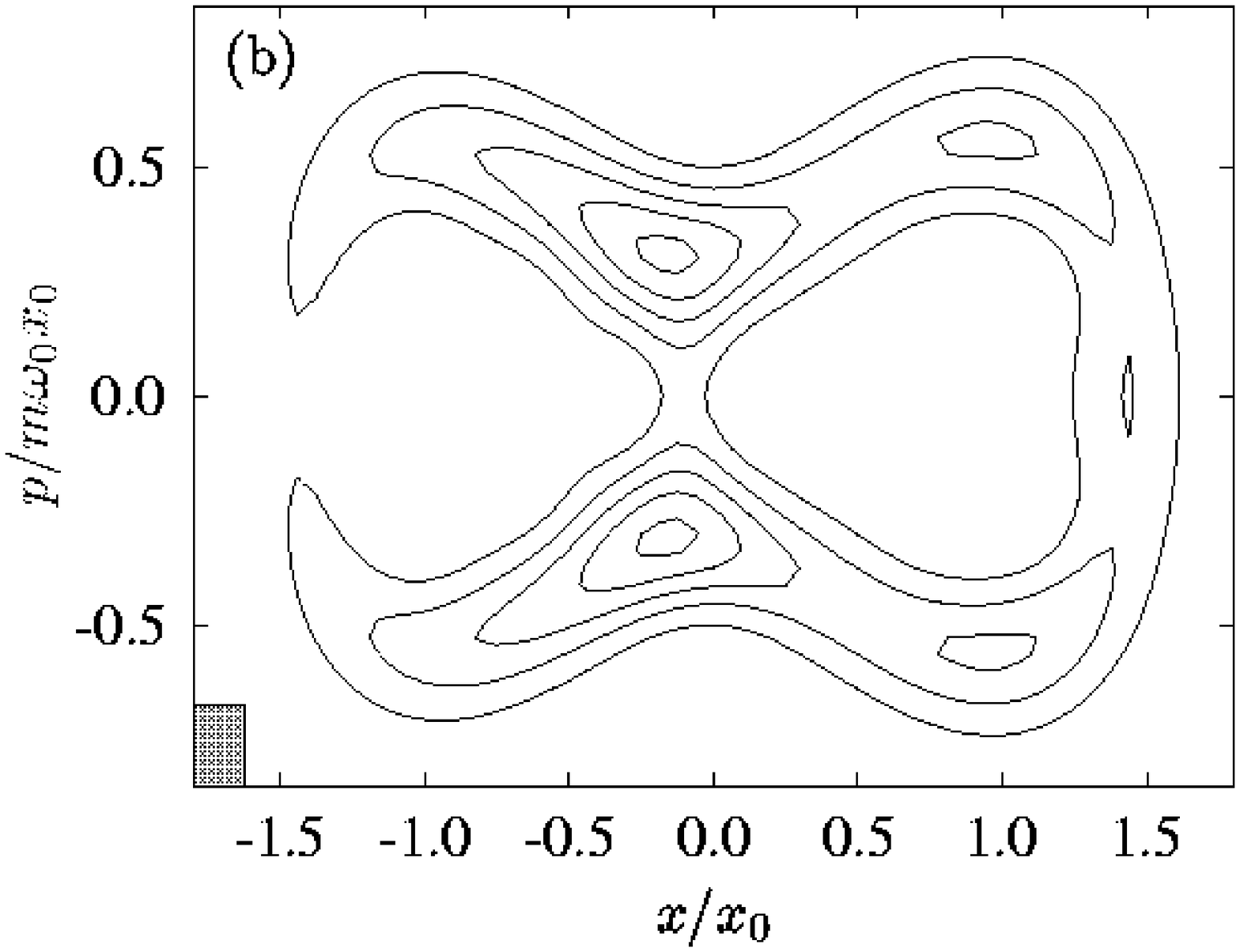}
}
\caption{ \label{ddw:fig:Qreg/chao}
Contour plots of the Husimi functions for the Floquet states
$|\phi_1^-\rangle\approx |\phi_{\rm r}^-\rangle$ (a) and
$|\phi_2^-\rangle\approx |\phi_{\rm c}^-\rangle$ (b)
of the harmonically driven quartic double well, Eq.\ (\protect\ref{ddw:H}),
at stroboscopic times $t=nP$.
The driving parameters $F=0.014$, $\Omega=0.982\,\omega_0$, are in
sufficient distance to the singlet-doublet crossing such that the admixture
from the chaotic singlet state is negligible.
The rectangle in the lower left corner has the size of the effective
quantum of action $\hbar_{\rm eff}$.
}
\end{figure}%

\begin{figure}[t]
\parbox[t]{7cm}{
\psfig{width=7cm,figure=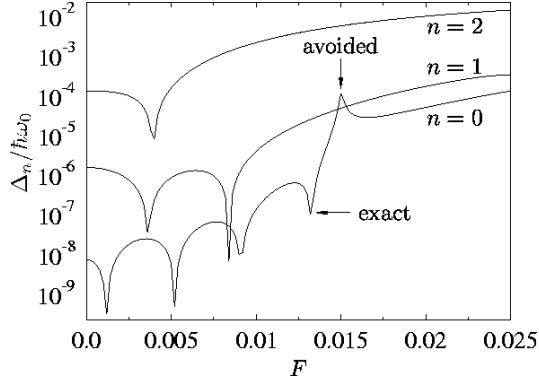}
}
\hfill\parbox[b]{4.5cm}{\caption{
Splitting of the lowest doublets for $D=4$ and $\Omega=0.982\,\omega_0$. The
arrows indicate the locations of the exact and the avoided crossing within a
three-level crossing of the type sketched in
Fig.\ \protect\ref{ddw:fig:crossing}a.
\label{ddw:fig:splitting} }\vspace{4ex}}
\end{figure}

In order to illustrate the above three-state model and to demonstrate its
adequacy, we have numerically studied a singlet-doublet crossing that occurs
for the double-well potential, Eq.~(\ref{ddw:H}), with $D=4$,
at a driving frequency $\Omega = 0.982\,\omega_0$ and amplitude $F=0.015029$
(Fig.~\ref{ddw:fig:num3s}).
The phase-space structure of the participating Floquet states
(Figs.~\ref{ddw:fig:poincare}, \ref{ddw:fig:Qreg/chao}) meets the
assumptions of our three-state theory.
A comparison of the appropriately scaled three-state theory
(Fig.~\ref{ddw:fig:theo3s}) with this real singlet-doublet crossing
(Fig.~\ref{ddw:fig:num3s}) shows satisfactory agreement. Note that in the
real crossing, the quasienergy of the chaotic singlet {\em decreases} as a
function of $F$, so that the exact crossing occurs to the left of the avoided
one.
This numerical example also shows a deficiency of the idealized three-state
model.
Following the global tendency of widening of the splittings with
increasing driving amplitude \cite{UtermannDittrichHanggi94,Wilkinson87,%
KohlerUtermannHanggiDittrich98}, it may happen that even far away from a
crossing, the doublet splitting does not return to its value on the
opposite side (see Fig.~\ref{ddw:fig:splitting}).
It is even possible that an exact crossing of $\epsilon_0^+$ and
$\epsilon_1^-$ never takes place in the vicinity of the crossing. In
that case, the relation of the quasienergies in the doublet gets reversed via
the crossing (Fig.~\ref{ddw:fig:crossing}c,d).
Nevertheless, the three-state scenario captures the essential features.

To study the dynamics of the tunneling process, we focus on the state
\begin{equation}
|\psi(t)\rangle = \frac{1}{\sqrt{2}}\left(
{\rm e}^{-{\rm i}\epsilon_0^+ t/\hbar}|\phi_0^+(t)\rangle
+{\rm e}^{-{\rm i}\epsilon_1^- t/\hbar}|\phi_1^-(t)\rangle\cos\beta
+{\rm e}^{-{\rm i}\epsilon_2^- t/\hbar}|\phi_2^-(t)\rangle\sin\beta
\right).
\label{eq:tunnelstate}
\end{equation}
It is constructed such that at $t = 0$, it corresponds to the decomposition of
$|\phi_{\rm R}\rangle$ in the basis (\ref{eq:psi012}) at finite distance
from the crossing. Therefore, it is initially localized in the regular region
in the right well and follows the time evolution under the Hamiltonian
(\ref{eq:ham3s}). From Eqs.~(\ref{eq:rightleft}), (\ref{eq:psi012}),
we find the probabilities for its evolving into $|\phi_{\rm R}\rangle$,
$|\phi_{\rm L}\rangle$, or $|\phi_{\rm c}\rangle$, respectively, to be
\begin{eqnarray}
P_{\rm R,L}(t)
&=& |\langle\phi_{\rm R,L}(t)|\psi(t)\rangle|^2 \nonumber\\
&=& \frac{1}{2}\Bigg(1 \pm \left[\cos\frac{(\epsilon_1^--\epsilon_0^+)t}{\hbar}
    \cos^2\beta+
\cos\frac{(\epsilon_2^--\epsilon_0^+)t}{\hbar}\sin^2\beta\right]
    \nonumber\\
&&  {} +\left[\cos\frac{(\epsilon_1^--\epsilon_2^-)t}{\hbar}-1\right]
    \cos^2\beta\sin^2\beta\Bigg), \label{eq:tun3s} \\
P_{\rm c}(t) &=& |\langle\phi_{\rm c}(t)|\psi(t)\rangle|^2 \nonumber
= \left[1-\cos\frac{(\epsilon_1^--\epsilon_2^-)t}{\hbar}\right]
    \cos^2\beta\sin^2\beta. \nonumber
\end{eqnarray}
We discuss the coherent dynamics of the three-state model for different
distances to the crossing and illustrate it by numerical results for the
real crossing introduced above.

\begin{figure}[t]
\centerline{
\psfig{width=6cm,figure=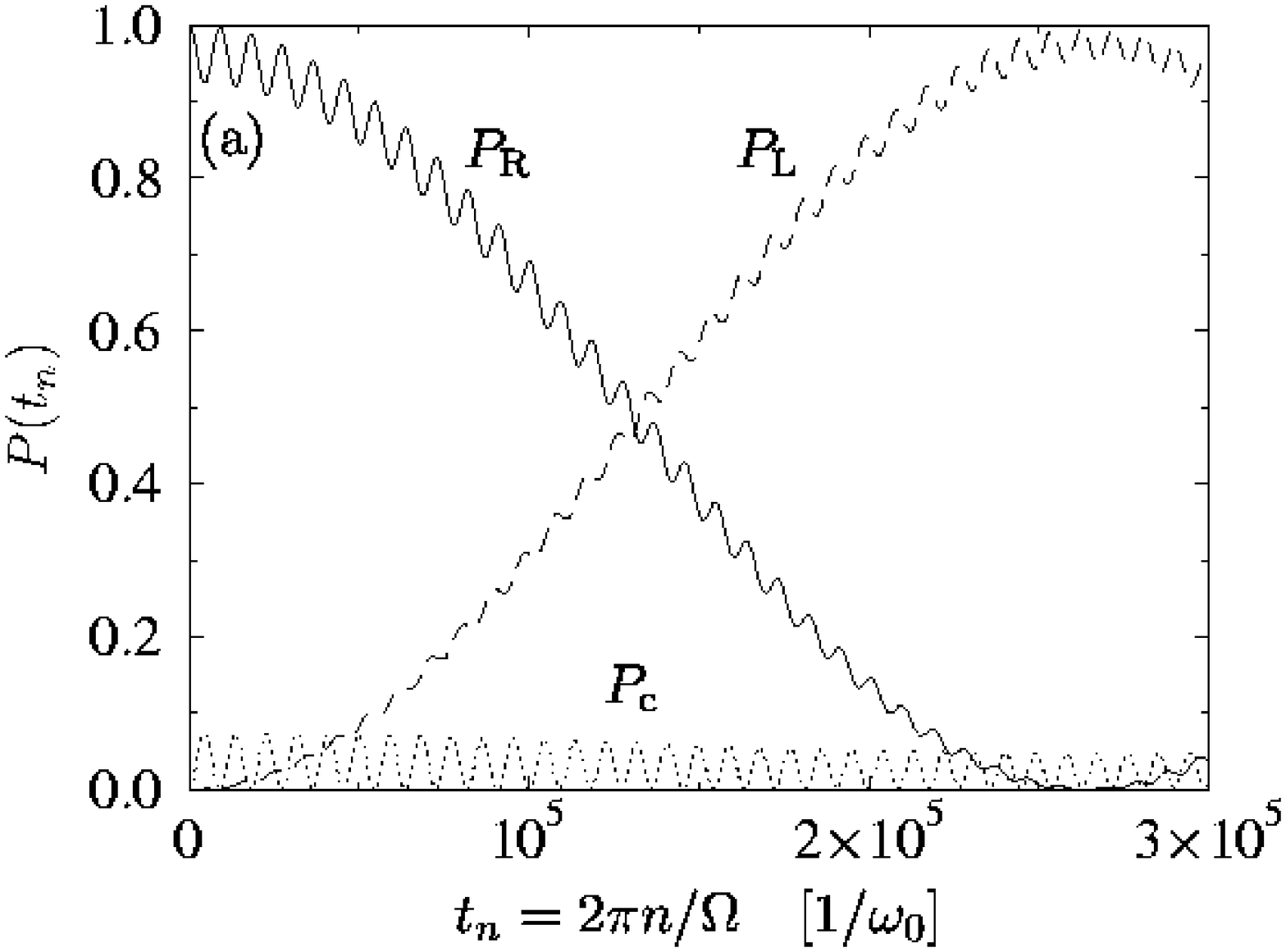}
\psfig{width=6cm,figure=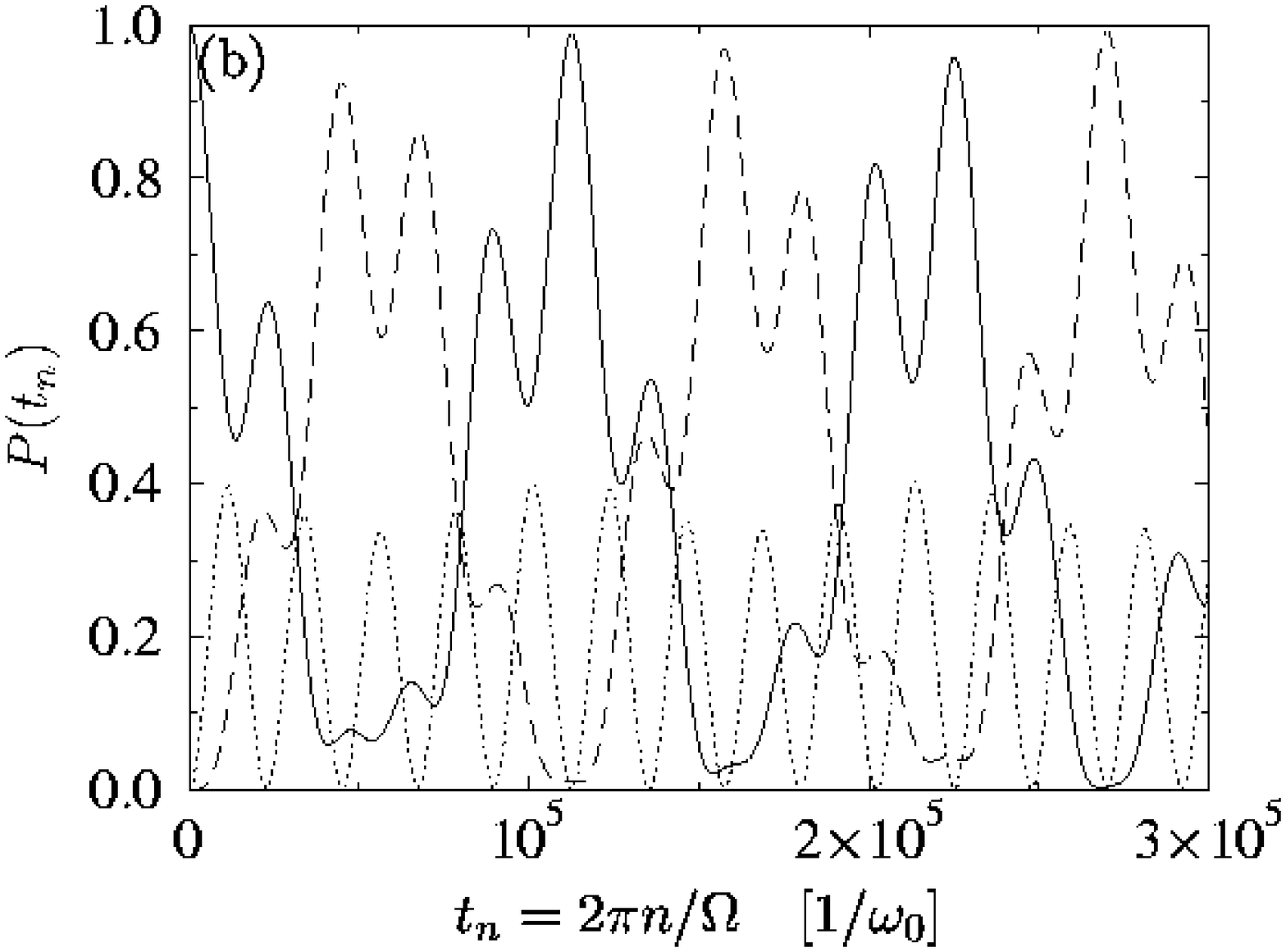}
}
\parbox[t]{6cm}{
\psfig{width=6cm,figure=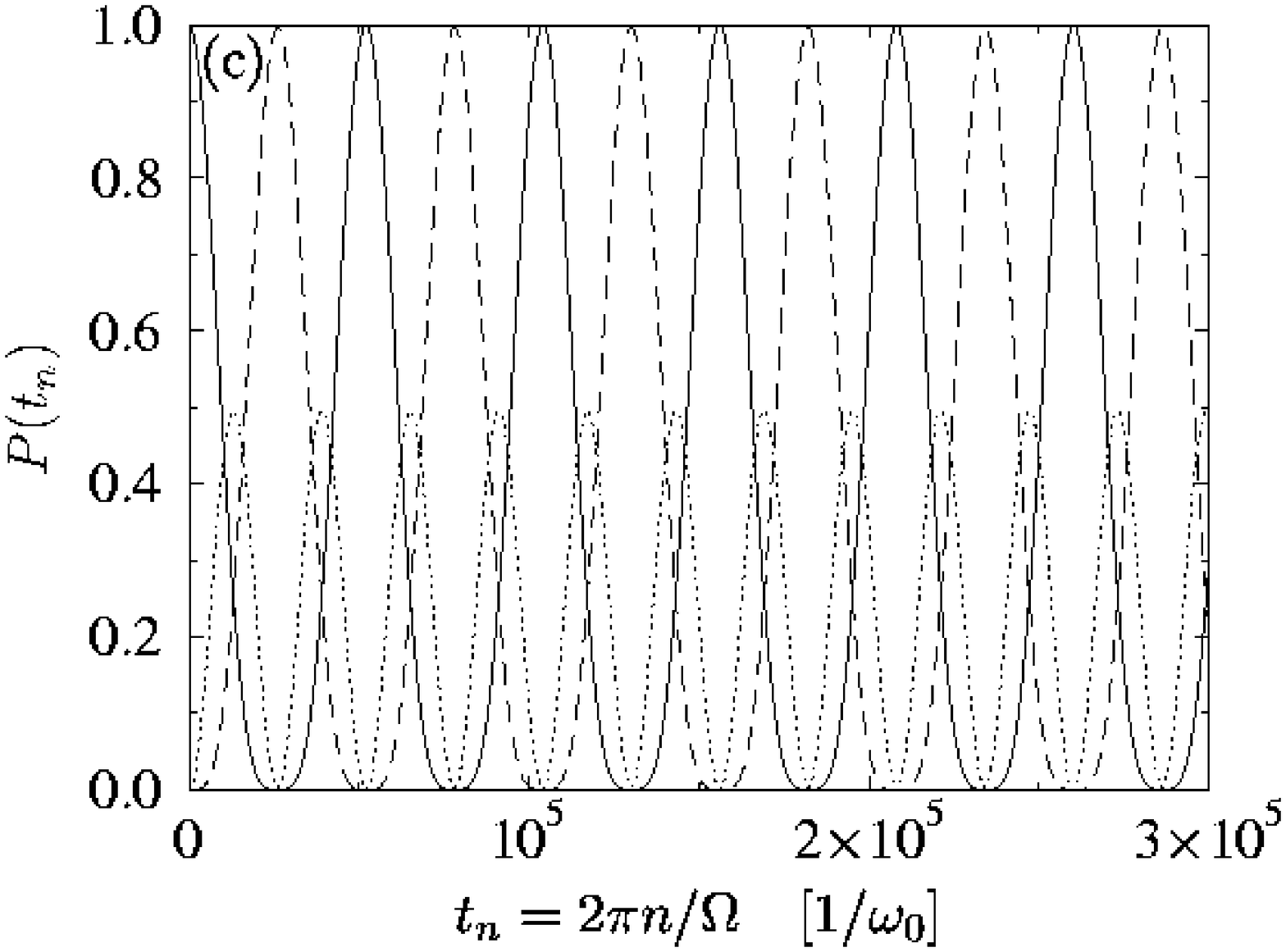}
}
\hfill\parbox[b]{5cm}{\caption{
Stroboscopic time evolution of a state initially localized in the right well,
in the vicinity of the singlet-doublet crossing shown in
Fig.\ \protect\ref{ddw:fig:num3s}, in terms of the probabilities to be in the
right well (which here is identical to the return probability, marked
by full lines), in the reflected
state in the left well (dashed), or in the chaotic state $|\psi_{\rm c}\rangle$
(dotted).  Parameter values are as in Fig.\ \protect\ref{ddw:fig:num3s}, and
$F = 0.0145$ (a), $0.0149$ (b), $0.015029$~(c).
\label{ddw:fig:tun3s}}}
\end{figure}

At sufficient distance from the crossing, there is only little mixing between
the regular and the chaotic states, i.e., $\sin\beta\ll 1$ or $\cos\beta\ll
1$. The tunneling process then follows the familiar two-state dynamics
involving only $|\phi_{\rm r}^+\rangle$ and $|\phi_{\rm r}^-\rangle$, with
tunnel frequency $\Delta/\hbar$ (Fig.~\ref{ddw:fig:tun3s}a). Close to the
avoided crossing, $\cos\beta$ and $\sin\beta$ are of the same order of
magnitude, and $|\phi_1^-\rangle$, $|\phi_2^-\rangle$ become very similar
to one
another. Both now have support in the chaotic layer as well as in the
symmetry-related regular regions, they are of a hybrid nature. Here, the
tunneling involves all the three states and must be described at least by a
three-level system. The exchange of probability between the two regular regions
proceeds via a ``stop-over'' in the chaotic region
\cite{BohigasTomsovicUllmo93,TomsovicUllmo94,LatkaGrigoliniWest94a,%
LatkaGrigoliniWest94b,LatkaGrigoliniWest94c}.

The three quasienergy differences that determine the time scales of this
process are in general all different, leading to complicated beats
(Fig.~\ref{ddw:fig:tun3s}b).
However, for $\Delta_{\rm c} = -2\Delta$, the two quasienergies
$\epsilon_1^- -\epsilon_0^+$ and $\epsilon_0^+ -\epsilon_2^-$ are degenerate.
At this point, the
center of the crossing, the number of different frequencies in the three-level
dynamics reduces to two again. This restores the familiar coherent tunneling
in the sense that there is again a simple periodic exchange of probability
between the regular regions
\cite{LatkaGrigoliniWest94a,LatkaGrigoliniWest94b,LatkaGrigoliniWest94c}.
However, the rate is much larger if
compared to the situation far off the crossing, and the chaotic region is now
temporarily populated during each probability transfer, twice per tunneling
cycle (Fig.~\ref{ddw:fig:tun3s}c).

\section{Incoherent quantum dynamics}
\label{sec:diss}
\subsection{Master equation}
\subsubsection{System-bath model.}
To achieve a microscopic model of dissipation, we couple the system
(\ref{ddw:H}) bilinearly to a bath of non-interacting harmonic
oscillators \cite{Magalinskii59,Zwanzig73}.
The total Hamiltonian of system and bath is then given by
\begin{equation}
H(t) = H_{\rm DW}(t)+
\sum_{\nu=1}^\infty \left(\frac{p_\nu^2}{2m_\nu}+\frac{m_\nu}{2}\omega_\nu^2
\left(x_\nu - \frac{g_\nu}{m_\nu\omega_\nu^2}x \right)^2 \right).
\end{equation}
The position $x$ of the system is coupled, with coupling strength $g_\nu$,
to an
ensemble of oscillators with masses $m_\nu$, frequencies $\omega_\nu$, momenta
$p_\nu$, and coordinates $x_\nu$. The bath is fully
characterized by the spectral density of the coupling energy,
\begin{equation}
\label{SpectralDensity}
J(\omega) = \pi\sum_{\nu=1}^\infty \frac{g_\nu^2}{2m_\nu\omega_\nu}
\delta(\omega-\omega_\nu).
\end{equation}

For the time evolution we choose an initial condition of the Feynman-Vernon
type: at $t=t_0$, the bath is in thermal equilibrium and uncorrelated
to the system, i.e.,
\begin{equation}
\label{FVinitial}
\rho(t_0) = \rho_{\rm S}(t_0)\otimes\rho_{\rm B,eq},
\end{equation}
where $\rho_{\rm B,eq}=\exp(-\beta H_{\rm B})/{\rm tr}_{\rm B}
\exp(-\beta H_{\rm B})$
is the canonical ensemble of the bath and $1/\beta = k_{\rm B}T$.

Due to the bilinearity of the system--bath coupling, one can always eliminate
the bath variables to get an exact, closed integro-differential equation for
the reduced density matrix $\rho_{\rm S}(t)={\rm tr}_{\rm B}\rho(t)$. It
describes the dynamics of the central system, subject to dissipation
\cite{Haake73}.

\subsubsection{Born-Markov approximation.}
\label{diss:markov}
%
In most cases, however, the integro-differen\-tial equation for $\rho_{\rm
S}(t)$ can be solved only approximately. In particular, in the limit of weak
coupling,
\begin{eqnarray}
\gamma & \ll & k_{\rm B}T/\hbar,\\
\gamma & \ll & |\epsilon_\alpha-\epsilon_{\alpha'}|/\hbar,
\label{mar:cond_frequency}
\end{eqnarray}
it is possible to truncate the time-dependent perturbation expansion in the
system--bath interaction after the second-order term. The quantity $\gamma$, to
be defined below, denotes the effective damping of the dissipative system, and
$|\epsilon_\alpha-\epsilon_{\alpha'}|/\hbar$ are the transition frequencies of
the central system. In the present case, {\em the central system is
understood to
include the driving} \cite{BlumelGrahamSirkoSmilanskyWaltherYamada89,%
BlumelBuchleitnerGrahamSirkoSmilanskyWalther91,GrahamHubner94,%
KohlerDittrichHanggi97}, so that the transition
frequencies are given by quasienergy differences. The autocorrelations of the
bath decay on a time scale $\hbar/k_{\rm B}T$ and thus in the present limit,
instantaneously on the time scale $1/\gamma$ of the system correlations.  With
the initial preparation (\ref{FVinitial}), the equation of motion for the
reduced density matrix in this approximation is given by
\cite{KohlerDittrichHanggi97}
\begin{eqnarray}
\nonumber
\dot \rho_{\rm S}(t)
&=& -\frac{{\rm i}}{\hbar}\left[ H_{\rm S}(t),\rho_{\rm S}(t) \right]
+\frac{1}{\pi\hbar}\int_{-\infty}^\infty {\rm d}\omega\, J(\omega)n_{\rm th}
(\hbar\omega) \nonumber\\
&&\times \int_0^\infty {\rm d}\tau\left( {\rm e}^{{\rm i}\omega\tau}
\left[ \tilde x(t-\tau,t)\rho_{\rm S}(t),x\right] + {\rm H.c.}\right),
\label{MasterEquationGeneral}
\end{eqnarray}
where $\tilde x(t',t)$ denotes the position operator in the interaction
picture defined by
\begin{equation}
\tilde x(t',t) = U^\dagger(t',t)\, x\, U(t',t),
\end{equation}
with $U(t',t)$, the propagator of the conservative driven double well, given
in Eq.~(\ref{eq:floquetprop}). `H.c.' means `Hermitian conjugate', and
\begin{equation}
n_{\rm th}(\epsilon)
=\frac{1}{{\rm e}^{\epsilon/k_{\rm B}T} - 1}=-n_{\rm th}(-\epsilon) - 1
\end{equation}
is the thermal occupation of the bath oscillator with energy $\epsilon$.  To
achieve a more compact notation, we require $J(-\omega)=-J(\omega)$.  In the
following, we shall restrict ourselves to an Ohmic bath, $J(\omega) =
m\gamma\omega$. This defines the effective damping constant $\gamma$.

We use the time-periodic components $|\phi_\alpha(t)\rangle$ of the Floquet
states as a basis to expand the density operator,
Eq.~(\ref{MasterEquationGeneral}). Expressing the matrix elements
\begin{equation}
X_{\alpha\beta}(t) = \langle\phi_\alpha(t)|x|\phi_\beta(t)\rangle
\end{equation}
of the position operator by their Fourier coefficients
\begin{eqnarray}
X_{\alpha\beta,n}
&=& \langle\langle\phi_\alpha(t)|x\,{\rm e}^{-{\rm i}n\Omega t}
|\phi_\beta(t)\rangle\rangle = X_{\beta\alpha,-n}^*\, , \\
X_{\alpha\beta}(t) &=& \sum_n {\rm e}^{{\rm i}n\Omega t}X_{\alpha\beta,n}\, ,
\end{eqnarray}
yields the equation of motion for the elements $\rho_{\alpha\beta}$ of the
reduced density matrix $\rho_{\rm S}$
\cite{DittrichOelschlagelHanggi93,DittrichHanggiOelschlagelUtermann95,%
BlumelBuchleitnerGrahamSirkoSmilanskyWalther91,KohlerDittrichHanggi97},
\begin{eqnarray}
\dot\rho_{\alpha\beta}(t)
&=& {{\rm d}\over{\rm d}t}\langle\phi_\alpha(t)|\rho_{\rm S}(t)|
    \phi_\beta(t)\rangle \nonumber\\
&=& -\frac{\rm i}{\hbar}(\epsilon_\alpha-\epsilon_\beta)
    \rho_{\alpha\beta}(t) \nonumber\\
&&  +\sum_{\alpha'\beta'nn'}\big( N_{\alpha\alpha',n}X_{\alpha\alpha',n}
    \rho_{\alpha'\beta'}X_{\beta'\beta,n'}  \nonumber\\
&&  {} -N_{\alpha'\beta',n}X_{\alpha\alpha',n'}X_{\alpha\alpha',n}
    \rho_{\beta'\beta}\big) {\rm e}^{{\rm i}(n+n')\omega t}
    + {\rm H.c.}\, .
\label{MasterEquationFloquet}
\end{eqnarray}
The coefficients of this differential equation are periodic
in time with the period of the driving.
The $N_{\alpha\beta,n}$ are given by
\begin{equation}
N_{\alpha\beta,n} = N(\epsilon_\alpha-\epsilon_\beta+n\hbar\Omega),\quad
N(\epsilon) = \frac{m\gamma\epsilon}{\hbar^2} n_{\rm th}(\epsilon).
\end{equation}
For $\epsilon\gg k_{\rm B}T$, $N(\epsilon)$ approaches zero.

Since the position operator $x$ is odd under ${\sf P}_P$ (cf.\
Eq.~(\ref{parity})), the master equations (\ref{MasterEquationGeneral})
and (\ref{MasterEquationFloquet}) are invariant under ${\sf P}_P$.
Therefore, both the conservative
and the dissipative dynamics preserve the parity of the operator
$|\phi_\alpha\rangle\langle\phi_\beta|$. If $|\phi_\alpha\rangle$ and
$|\phi_\beta\rangle$ belong to the same parity class, it is even, and odd
otherwise. In particular, the projectors
$|\phi_\alpha\rangle\langle\phi_\alpha|$ and thus all density matrices diagonal
in the Floquet basis are even under ${\sf P}_P$.

\subsubsection{Rotating-wave approximation.}
%
Assuming that dissipative effects are relevant only on a time scale much
larger than the period $P$ of the driving, we average the
coefficients of the master equation (\ref{MasterEquationFloquet}) over
$P$ to obtain the equation of motion
\begin{equation}
\dot\rho_{\alpha\beta}(t)
=-\frac{\rm i}{\hbar}(\epsilon_\alpha-\epsilon_\beta)\rho_{\alpha\beta}(t)
+\sum_{\alpha'\beta'} {\cal L}_{\alpha\beta,\alpha'\beta'}\rho_{\alpha'\beta'},
\label{mastereq}
\end{equation}
with the time-independent dissipative part
\begin{eqnarray}
{\cal L}_{\alpha\beta,\alpha'\beta'}
&=& \sum_n \left( N_{\alpha\alpha',n} + N_{\beta\beta',n}\right)
    X_{\alpha\alpha',n} X_{\beta'\beta,-n}
\nonumber\\
&& -\delta_{\beta\beta'}\sum_{\beta'',n}
   N_{\beta''\alpha',n}X_{\alpha\beta'',-n}
   X_{\beta''\alpha',n}\nonumber\\
&& -\delta_{\alpha\alpha'}\sum_{\alpha''n}
   N_{\alpha''\beta',n}
   X_{\beta'\alpha'',-n}X_{\alpha''\beta,n} .
\label{MasterEquationRWA}
\end{eqnarray}
This step amounts to a rotating-wave approximation which is,
however, less restrictive than the one introduced in
\cite{BlumelGrahamSirkoSmilanskyWaltherYamada89,%
BlumelBuchleitnerGrahamSirkoSmilanskyWalther91}
where dissipative effects are averaged over the generally
longer time scale ${\rm max}_{\alpha,\beta,n}(2\pi\hbar
/(\epsilon_\alpha-\epsilon_\beta+n\hbar\Omega))$.

\subsection{Chaos-assisted dissipative tunneling}
\label{ddw:cat:diss}

The crucial effect of dissipation on a quantum system is the disruption of
coherence: a coherent superposition evolves into an incoherent mixture.  Thus,
phenomena based on coherence, such as tunneling, are rendered transients that
fade out on a finite time scale $t_{\rm decoh}$. In general, for driven
tunneling in the weakly damped regime, this time scale gets shorter for higher
temperatures, as transition rates grow \cite{HanggiTalknerBorkovec90}. However,
in the vicinity of an exact crossing of the ground-state quasienergies, the
coherent suppression of tunneling
\cite{GrossmannDittrichJungHanggi91,GrossmannJungDittrichHanggi91,%
GrifoniHanggi98} can be stabilized with higher temperatures
\cite{DittrichOelschlagelHanggi93,OelschlagelDittrichHanggi93,%
DittrichHanggiOelschlagelUtermann95} and increasing friction
\cite{MakarovMakri95,Makri97} until levels outside the doublet start to play a
r\^ole. We have studied dissipative chaos-assisted tunneling, at
the particular real singlet-doublet crossing introduced in Sec.~\ref{sec:3s}
(see Fig.~\ref{ddw:fig:num3s}).
The time evolution has been computed numerically by
iterating the dissipative quantum map for the improved master equation in
moderate rotating-wave approximation, Eq.~(\ref{mastereq}).
As initial condition, we have chosen the density operator
$\rho(0)=|\phi_{\rm R}\rangle\langle\phi_{\rm R}|$, a pure state located
in the right well.

\begin{figure}[t]
\centerline{
\psfig{width=6cm,figure=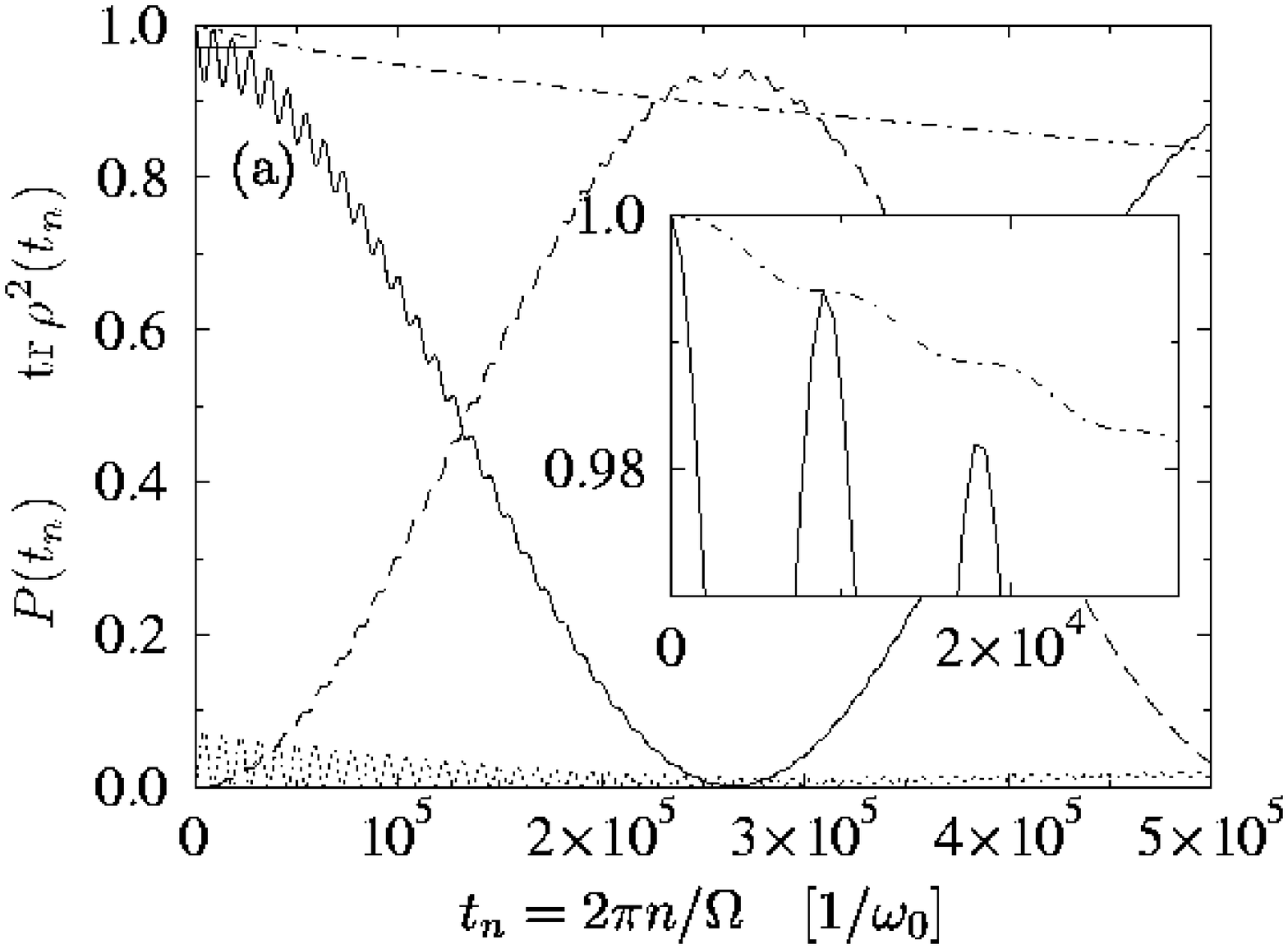}
\psfig{width=6cm,figure=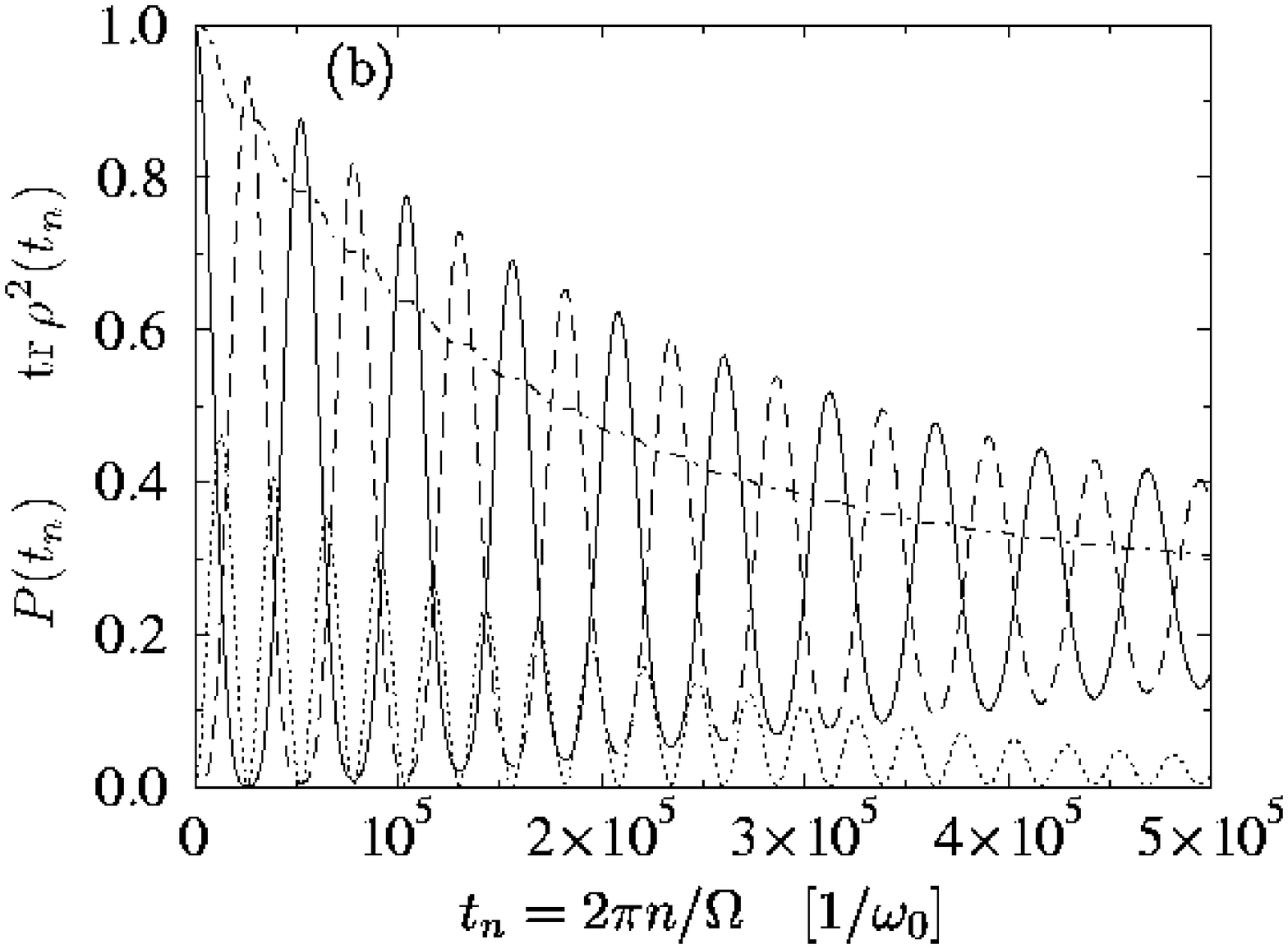}
}
\caption{
Occupation pro\-ba\-bilities as in Fig.\ \protect\ref{ddw:fig:tun3s}a,c,
but in the presence of dissipation.
The dash-dotted line shows the time evolution of
${\rm tr}\,\rho^2$. The parameter values are $D=4$, $\Omega=0.982\,\omega_0$,
$\gamma=10^{-6}\omega_0$, $k_{\rm B}T=10^{-4}\hbar\omega_0$, and $F=0.0145$
(a), $0.015029$ (b). The inset in (a) is a blow up of the rectangle in the
upper left corner of that panel. \label{ddw:fig:short}
}
\parbox[t]{6cm}{
\psfig{width=6cm,figure=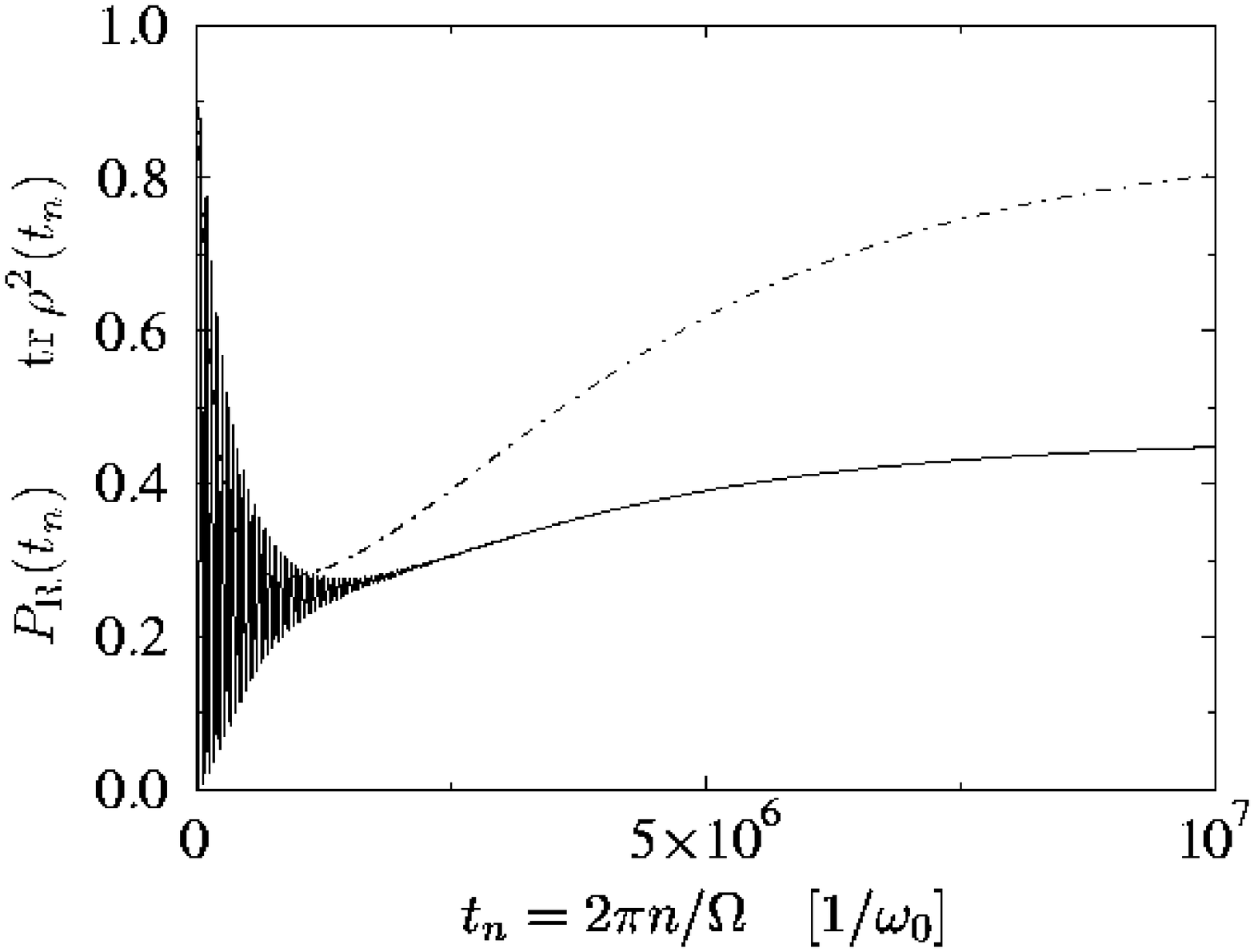}
}
\hfill\parbox[b]{5.2cm}{\caption{
Time evolution of the return probability $P_{\rm R}$ (full line) and the
coherence function ${\rm tr}\,\rho^2$ (dash-dotted) during loss and regain
of coherence. The parameter values are as in Fig.~\protect\ref{ddw:fig:short}b.
\label{ddw:fig:recoherence} }\vspace{2ex}}
\end{figure}

In the vicinity of a singlet-doublet crossing, the tunnel splitting
increases signifi\-cantly---the essence of chaos-assisted tunneling. During
the tunneling, the chaotic singlet $|\phi_{\rm c}\rangle$ becomes populated
periodically with frequency $|\epsilon_2^--\epsilon_1^-|/\hbar$, cf.\
Eq.~(\ref{eq:tun3s}) and Fig.~\ref{ddw:fig:tun3s}. The high mean energy of this
singlet results in an enhanced decay of coherence at times when it is well
populated (Fig.~\ref{ddw:fig:short}). For the relaxation towards the asymptotic
state, also the slower transitions within doublets are relevant. Therefore, the
corresponding time scale $t_{\rm relax}$ can be much larger than  $t_{\rm
decoh}$ (Fig.~\ref{ddw:fig:recoherence}).

To obtain quantitative estimates for the dissipative time scales, we
approximate $t_{\rm decoh}$ by the decay rate of ${\rm tr}\,\rho^2$, as a
measure of coherence, averaged over a time $t_{\rm p}$,
\begin{eqnarray}
\frac{1}{t_{\rm decoh}}
&=& -\frac{1}{t_{\rm p}}\int_0^{t_{\rm p}} {\rm d}t' \frac{\rm d}{{\rm d}t'}\,
    {\rm tr}\,\rho^2(t') \\
&=& \left.\frac{1}{t_{\rm p}}\Big( {\rm tr}\,\rho^2(0)
- {\rm tr}\,\rho^2(t_{\rm p})\Big)
\right. .
\end{eqnarray}
Because of the stepwise loss of coherence (Fig.~\ref{ddw:fig:short}),
we have chosen the propagation time $t_{\rm p}$ as an $n$fold multiple of the
duration $2\pi\hbar/|\epsilon_2^- - \epsilon_1^-|$ of the chaotic beats.
For this procedure to be meaningful, $n$ should be so large that the coherence
decays substantially during the time $t_{\rm p}$ (in our numerical studies to
a value of approximately 0.9).
The time scale $t_{\rm relax}$ of the approach to the asymptotic state is
given by the reciprocal of the smallest real part of the eigenvalues of
the dissipative kernel.

\begin{figure}[t]
\centerline{
\psfig{width=6cm,figure=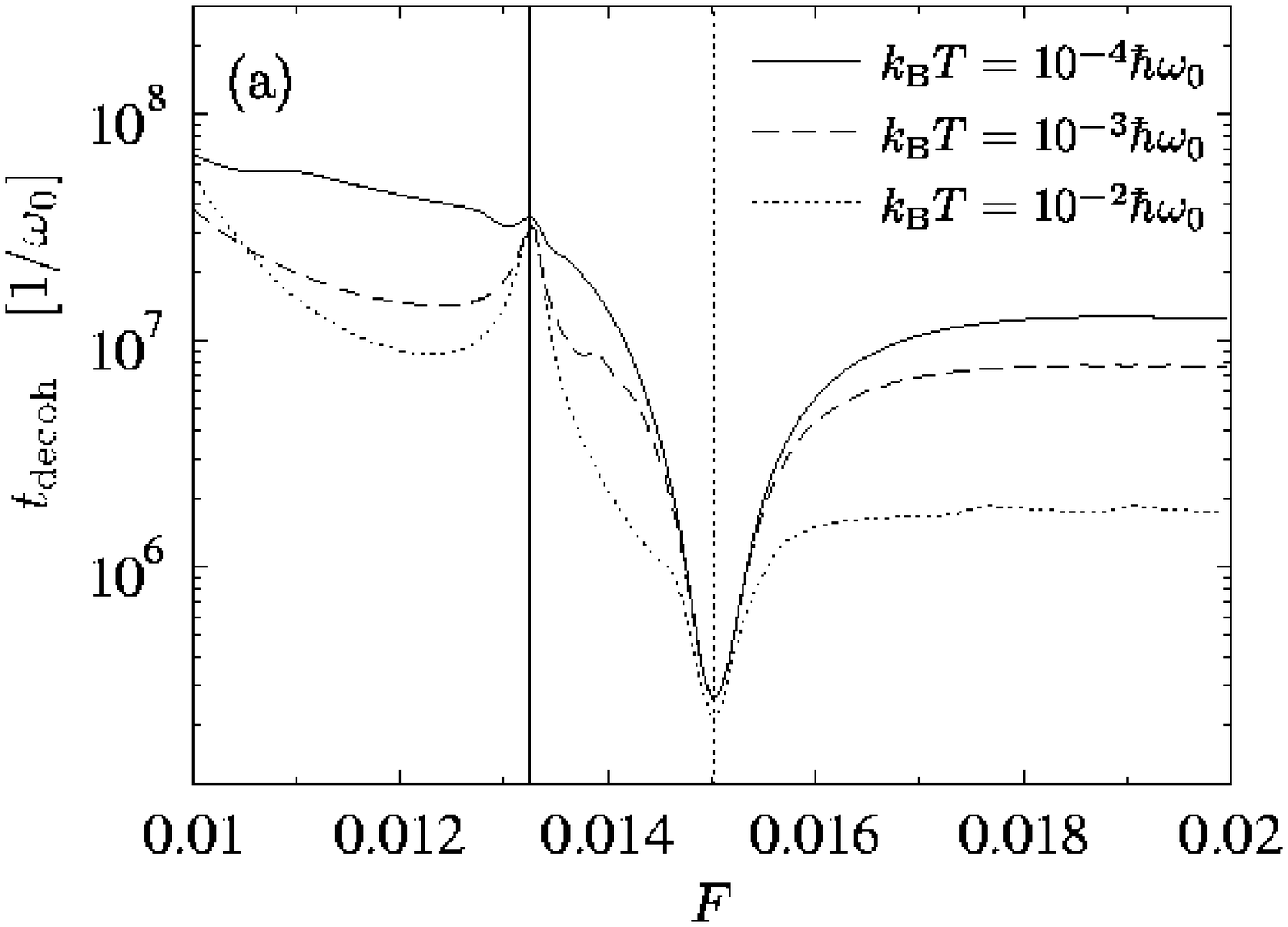}
\psfig{width=6cm,figure=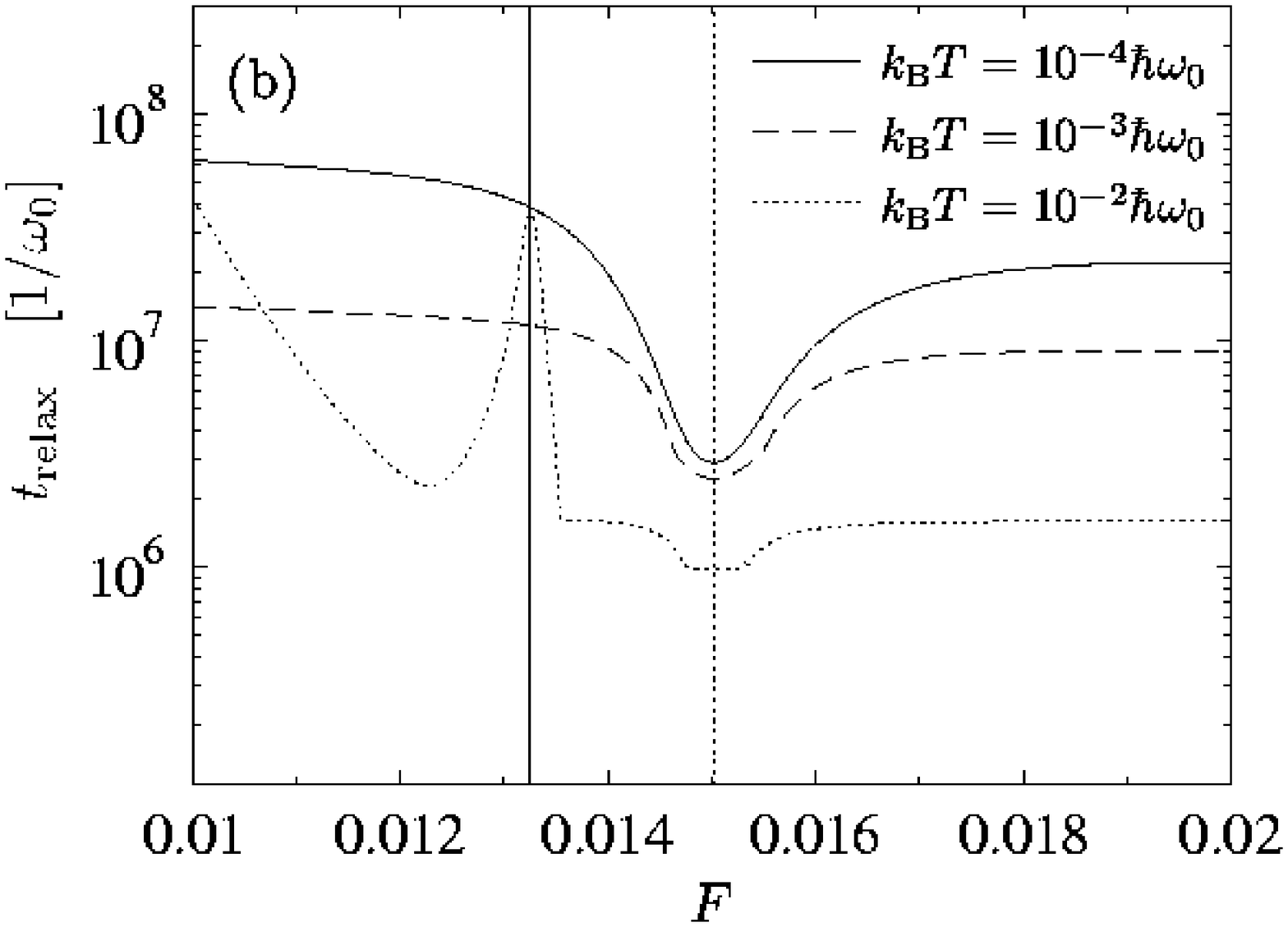}
}
\caption{
Time scales of the decay of the coherence measure ${\rm tr}\,\rho^2$ (a)
and of the relaxation
towards the asymptotic solution (b) near the singlet-doub\-let crossing. Near
the exact crossing ($F\approx 0.013$, full vertical line) coherence is
stabilized, whereas at the center of the avoided crossing ($F\approx 0.015$,
dashed vertical line) the decay of coherence is accelerated. The parameter
values are $D=4$, $\Omega=0.982\,\omega_0$, $\gamma=10^{-6}\omega_0$,
temperature as given in the legend.  \label{ddw:fig:timescales} }
\end{figure}

Outside the singlet-doublet crossing we find that the decay of coherence and
the relaxation take place on roughly the same time scale
(Fig.~\ref{ddw:fig:timescales}). At $F \approx 0.013$, the chaotic singlet
induces an exact crossing of the ground-state quasienergies
(see Fig.~\ref{ddw:fig:splitting}),
resulting in a {\it stabilization of coherence\/} with
increasing temperature. At the center of the avoided crossing, the decay of
coherence becomes much faster and is essentially independent of
temperature. This indicates that transitions from states with mean energy far
above the ground state play a crucial r\^ole.

\subsection{Asymptotic state}
\label{ddw:cat:attractor}

As the dynamics described by the master equation (\ref{mastereq})
is dissipative, it converges in the long-time limit to an asymptotic state
$\rho_\infty(t)$. In general, this attractor remains time dependent but shares
 the symmetries of the central system, i.e.\ here, periodicity and
generalized parity. However, the coefficients
(\ref{MasterEquationRWA}) of the master equation for the matrix elements
$\rho_{\alpha\beta}$, valid within a moderate rotating-wave approximation, are
time independent and so the asymptotic solution also is.
The explicit time dependence of the attractor has been effectively
eliminated by
representing it in the Floquet basis and introducing a mild rotating-wave
approximation.

To gain some qualitative insight into the asymptotic solution, we focus on
the diagonal elements
\begin{equation}
{\cal L}_{\alpha\alpha,\alpha'\alpha'}
=2\sum_n N_{\alpha\alpha',n}|X_{\alpha\alpha',n}|^2,\quad \alpha\neq\alpha',
\label{fullRWA}
\end{equation}
of the dissipative kernel. They give the rates of direct transitions from
$|\phi_{\alpha'}\rangle$ to $|\phi_\alpha\rangle$. Within a full
rotating-wave approximation \cite{BlumelGrahamSirkoSmilanskyWaltherYamada89,%
BlumelBuchleitnerGrahamSirkoSmilanskyWalther91}, these were the only
non-vanishing contributions to the master equation to affect the
diagonal elements $\rho_{\alpha\alpha}$ of the density matrix.

In the case of zero driving amplitude, the Floquet states
$|\phi_\alpha\rangle$ reduce to the eigenstates of the undriven Hamiltonian
$H_{\rm DW}$. The only non-vanishing Fourier component is then
$|c_{\alpha,0}\rangle$, and the quasienergies $\epsilon_\alpha$ reduce to the
corresponding eigenenergies $E_\alpha$.
Thus ${\cal L}_{\alpha\alpha,\alpha'\alpha'}$ only consists of a single term
proportional to $N(\epsilon_\alpha-\epsilon_{\alpha'})$.
It describes two kinds of thermal transitions:
decay to states with lower energy and, if the energy difference is less than
$k_{\rm B}T$, thermal activation to states with higher energy. The ratio of
the direct transitions forth and back then reads
\begin{equation}
{{\cal L}_{\alpha\alpha,\alpha'\alpha'}\over{\cal L}_{\alpha'\alpha'
,\alpha\alpha}}
=\exp\left(-{\epsilon_\alpha-\epsilon_{\alpha'} \over k_{\rm B}T}\right).
\end{equation}
We have detailed balance and therefore the steady-state solution is
\begin{equation}
\rho_{\alpha\alpha'}(\infty) \sim {\rm e}^{-\epsilon_\alpha/k_{\rm B}T}\,
\delta_{\alpha\alpha'}.
\end{equation}
In particular, the occupation probability decays
monotonically with the energy of the eigenstates. In the limit $k_{\rm
B}T\to0$,
the system tends to occupy the ground state only.

For a strong driving, each Floquet state $|\phi_\alpha\rangle$ contains a large
number of Fourier components and ${\cal L}_{\alpha\alpha,\alpha'\alpha'}$ is
given by a sum over contributions with quasienergies $\epsilon_\alpha -
\epsilon_{\alpha'} + n\hbar\Omega$. Thus decay to states with ``higher''
quasienergy (recall that quasienergies do not allow for a global
ordering) becomes possible due to terms with $n<0$. Physically, it amounts to
an incoherent transition under absorption of driving-field quanta.
Correspondingly, the system tends to occupy Floquet states comprising many
Fourier components with low index $n$. According to
Eq.~(\ref{eq:meanen:fourier}), these states have a low mean energy.

\begin{figure}[t]
\centerline{
\psfig{width=6cm,figure=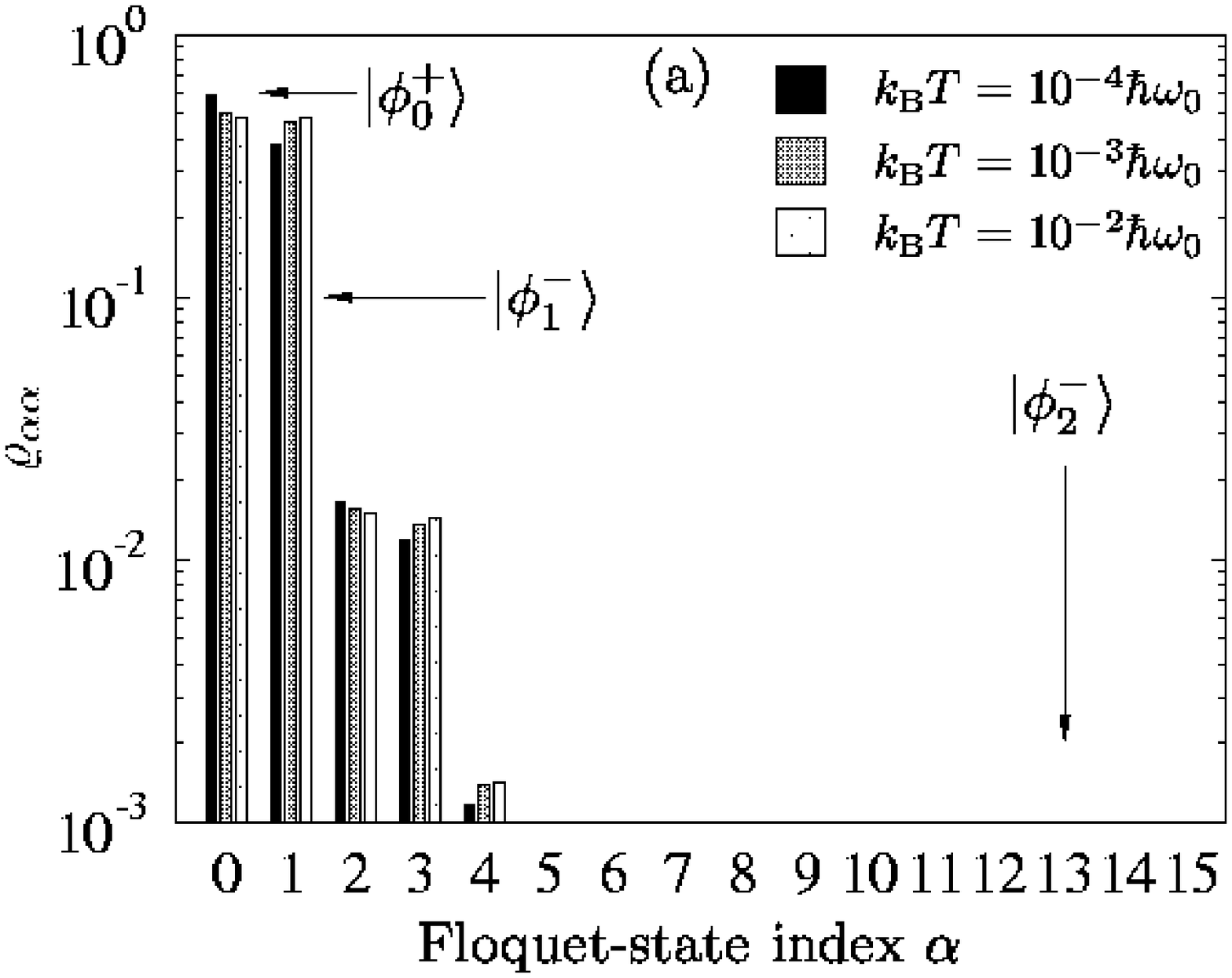}
\psfig{width=6cm,figure=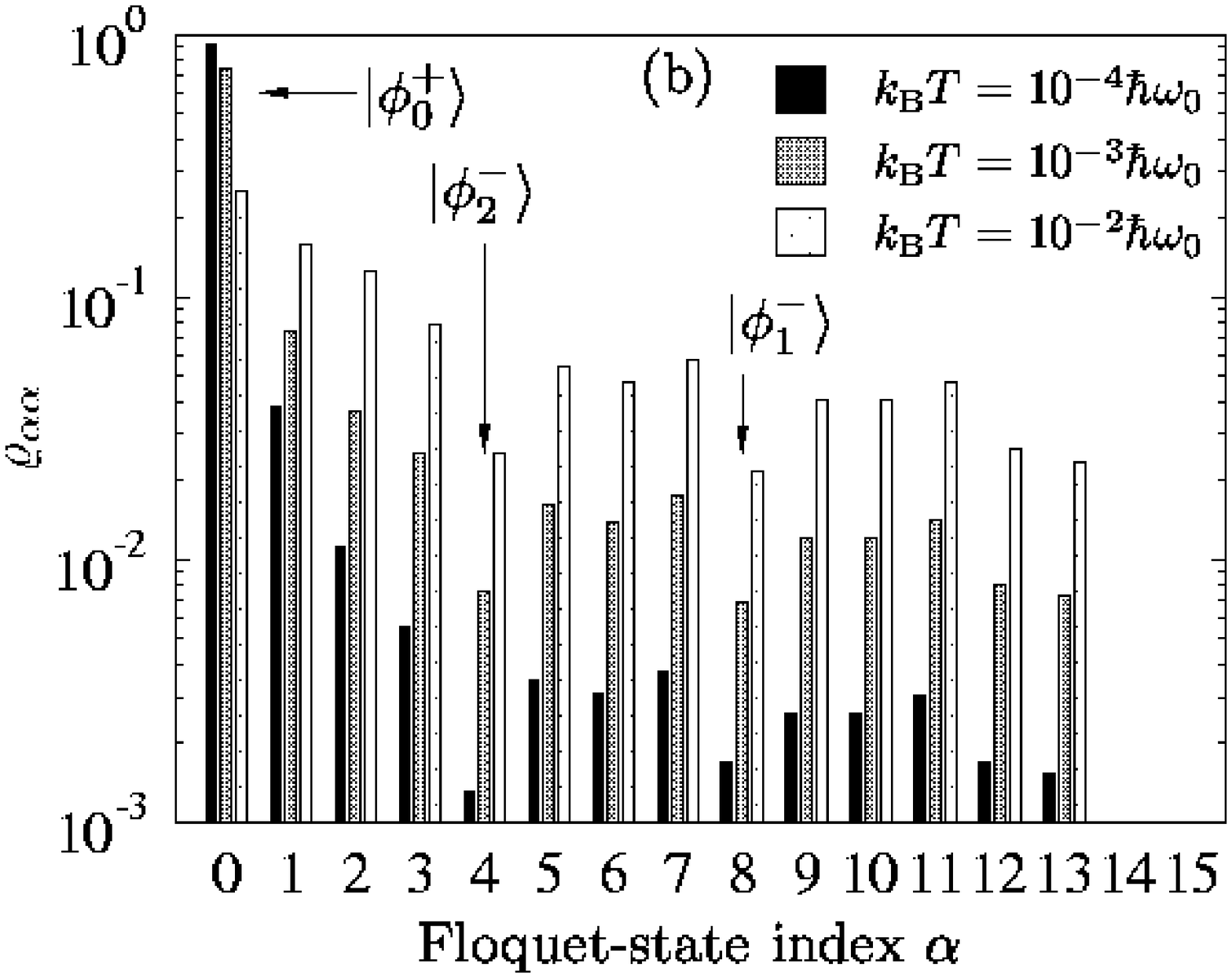}
}
\caption{
Occupation probability $\rho_{\alpha\alpha}$ of the Floquet states
$|\phi_\alpha\rangle$ in the long-time limit.
The parameter values are $D=4$, $\Omega=0.982\,\omega_0$,
$\gamma=10^{-6}\omega_0$, and $F = 0.013$ (a), $0.015029$ (b),
temperature as given in the legend. }
\label{ddw:fig:occupation}
\end{figure}

The effects under study are found for a driving with a frequency of the order
of unity. Thus for a quasienergy doublet, not close to a
crossing, we have $|\epsilon_\alpha-\epsilon_{\alpha'}| \ll \hbar\Omega$, and
${\cal L}_{\alpha'\alpha',\alpha\alpha}$ is dominated by contributions with
$n<0$, where the splitting has no significant influence.
However, up to the tunnel splitting, the two partners
in the quasienergy doublet are almost identical. Therefore, with respect to
dissipation, both should behave similarly. In particular, one expects an equal
population of the doublets even in the limit of zero temperature
(Fig.~\ref{ddw:fig:occupation}a), in contrast to the time-independent case.

In the vicinity of a singlet-doublet crossing the situation is more subtle.
Here, the odd partner, say, of the doublet mixes with a chaotic singlet, cf.\
Eq.~(\ref{eq:psi012}), and thus acquires components with higher energy.  Due
to the high mean energy $E_{\rm c}^-$ of the chaotic singlet, close to the top
of the barrier, the decay back to the ground state can also proceed indirectly
via other states with mean energy below $E_{\rm c}^-$.  Thus
$|\phi_1^-\rangle$ and $|\phi_2^-\rangle$ are depleted and mainly
$|\phi_0^+\rangle$ will be populated.  However, if the temperature is
significantly above the splitting $2b$ at the avoided crossing, thermal
activation
from $|\phi_0^+\rangle$ to $|\phi_{1,2}^-\rangle$, accompanied by depletion
via the states below $E_{\rm c}^-$, becomes possible.  Thus asymptotically,
all these states become populated in a steady closed flow
(Fig.~\ref{ddw:fig:occupation}b).
The long-time limit of the corresponding classical dynamics converges to one
of two limit cycles, each of which is located close to one of the potential
minima. In a stroboscopic map they correspond to two isolated fixed points.
This behavior is qualitatively different from the asymptotic limit of
the dissipative quantum dynamics near the center of the crossing and shows
that the occupation of levels outside the singlet and the doublet for $t \to
\infty$ is a pure quantum effect.
\begin{figure}[t]
\centerline{
\psfig{width=6cm,figure=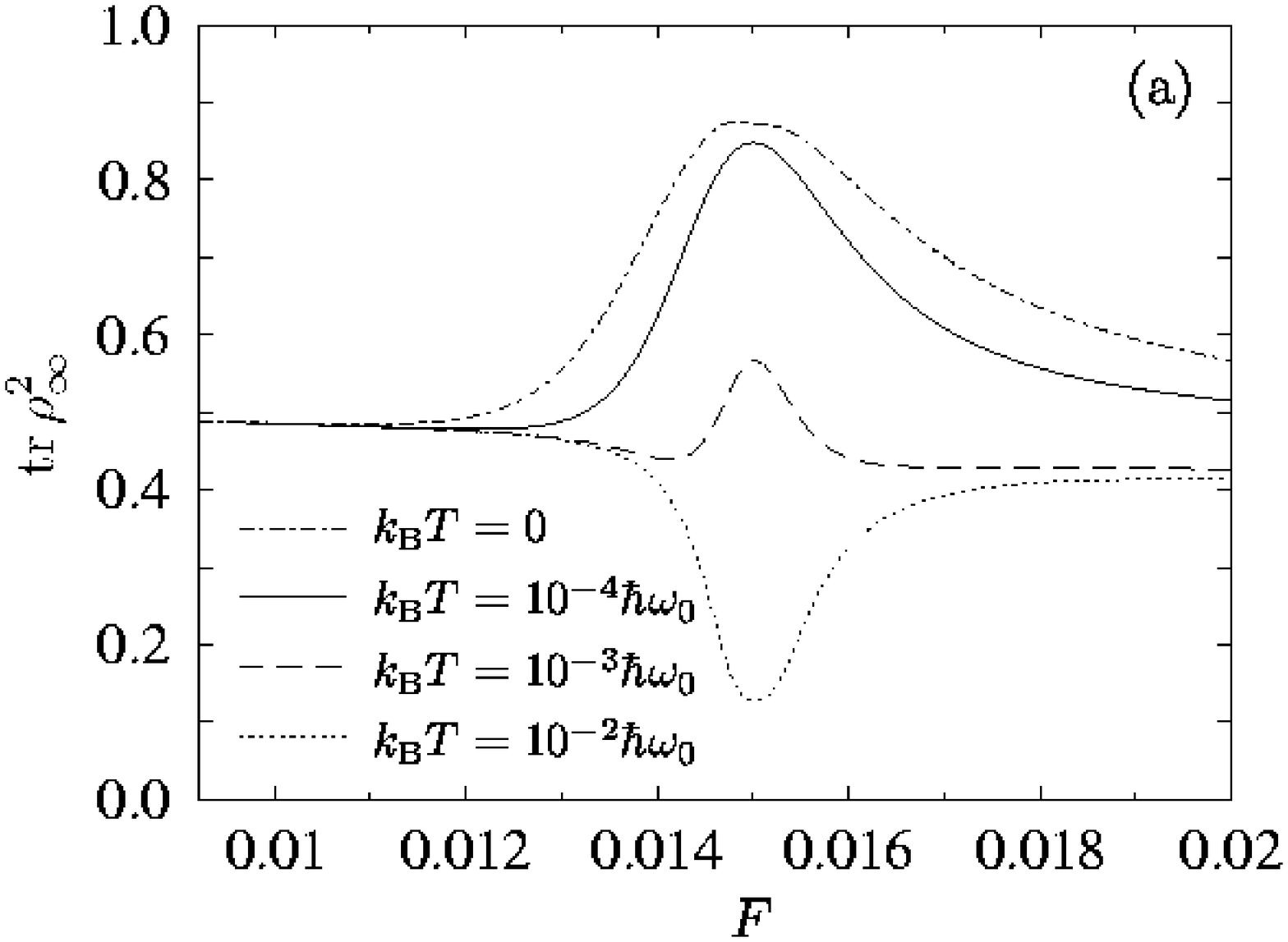}
\psfig{width=6cm,figure=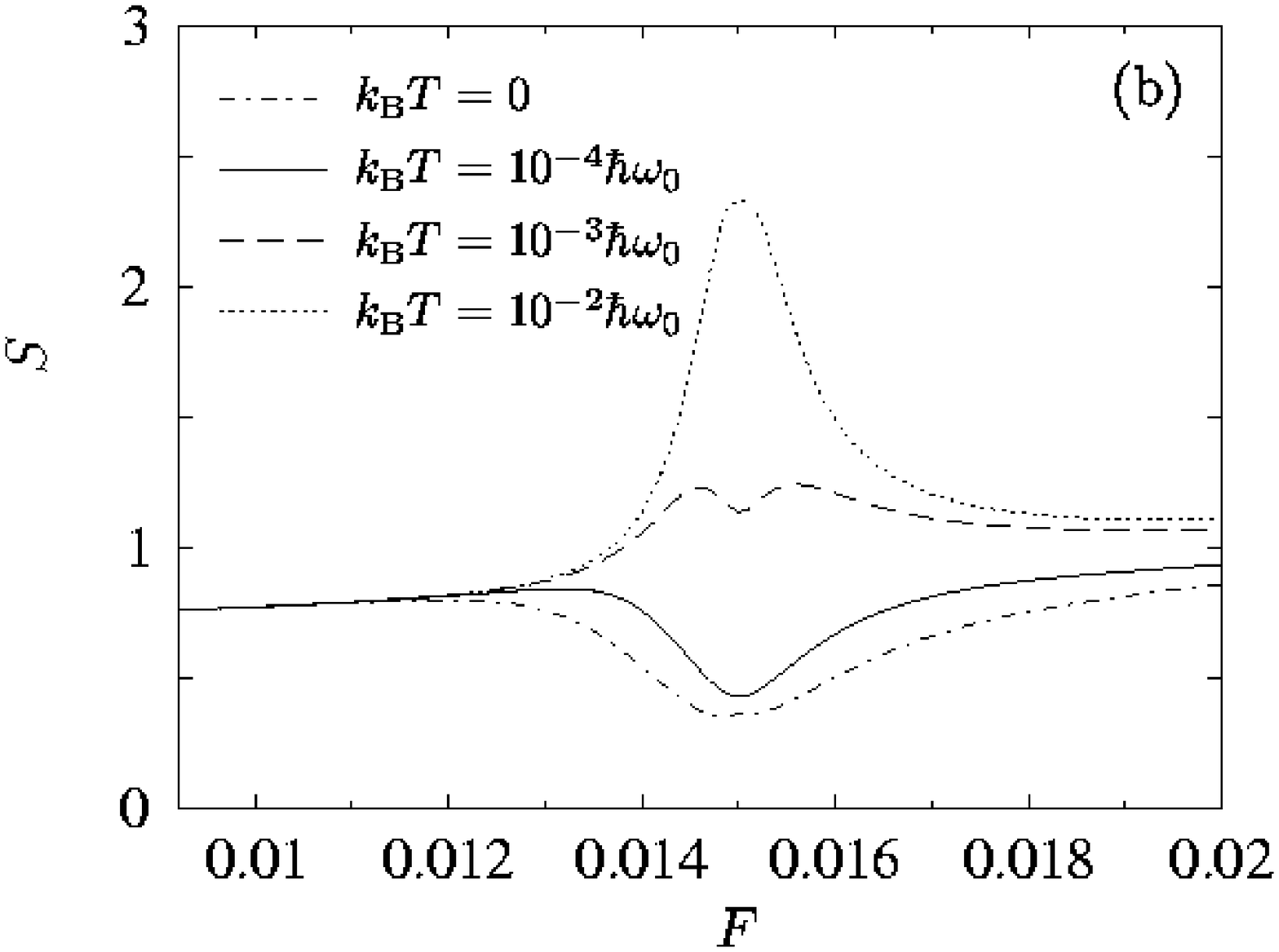}
}
\caption{ \label{ddw:fig:cic}
Coherence (a) and Shannon entropy (b) of the asymptotic state in the vicinity
of a singlet-doublet crossing for different temperatures as given in the
legend. The other parameter values are $D=4$, $\Omega = 0.982\,\omega_0$, and
$\gamma=10^{-6}\omega_0$.
}
\end{figure}
\begin{figure}[t]
\parbox[t]{6cm}{
\psfig{width=6cm,figure=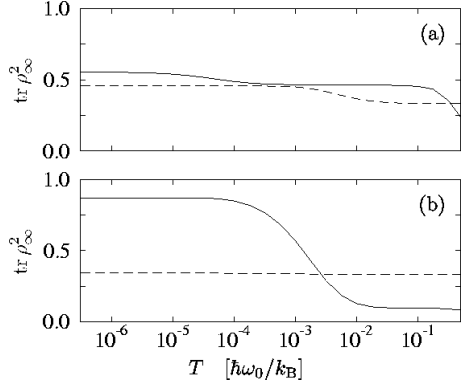}
}
\hfill\parbox[b]{5cm}{\caption{ \label{ddw:fig:coh}
Coherence of the asymptotic state in the vicinity of a singlet-doublet
crossing for $F=0.013$ (a) and $F=0.015029$ (b): exact calculation (full line)
compared to the result of a three-level description (dashed) of
the dissipative dynamics. The other parameter values are $D=4$,
$\Omega = 0.982\,\omega_0$, and $\gamma=10^{-6}\omega_0$.
}\vspace{2ex}}
\end{figure}%

Important global characteristics of the asymptotic state, measuring its degree
of spreading over phase space, are the Shannon entropy $S=-{\rm
tr}\,(\rho_\infty\ln\rho_\infty)$ or, alternatively, ${\rm tr}\,\rho_\infty^2$.
The latter gives approximately the inverse number of incoherently occupied
states and can be considered an ``incoherent inverse participation ratio''
\cite{DittrichSmilansky91}. It equals unity only if the attractor is a pure
state. According to the above scenario, we expect ${\rm tr}\,\rho_\infty^2$ to
assume the value $1/2$, in a regime with strong driving but preserved doublet
structure, reflecting the incoherent population of the ground-state doublet. In
the vicinity of the singlet-doublet crossing where the doublet structure is
dissolved, its value should be close to unity for temperatures $k_{\rm B}T\ll
2b$ and much less than unity for $k_{\rm B}T\gg 2b$ (Figs.~\ref{ddw:fig:cic}a,
\ref{ddw:fig:coh}). This means that the crossing of the chaotic singlet
with the
regular doublet leads to an improvement of coherence if the temperature is
below
the splitting of the avoided crossing, and to a loss for temperatures
above the splitting. This phenomenon amounts to a {\it chaos-induced}
coherence or
incoherence, respectively. The corresponding Shannon entropy (Fig.\
\ref{ddw:fig:cic}b) assumes approximately the value $\ln n$ for $n$
incoherently populated states. Thus outside the crossing, we have $S\approx\ln
2$ and at the center of the crossing the entropy exhibits a significant
temperature dependence.

The crucial r\^ole of the decay via states not involved in the three-level
crossing can be demonstrated by comparing it to the dissipative dynamics
including only these three levels (plus the bath). At the crossing, the
three-state model results in a completely different type of asymptotic state
(Fig.~\ref{ddw:fig:coh}).
The failure of the three-state model in the presence of
dissipation clearly indicates that in the vicinity of the singlet-doublet
crossing, it is important to take a large set of levels into account.

\section{Signatures of chaos in the asymptotic state}
\label{ddw:sec:attractor}

Phase-space representations
of quantum mechanics, like the Husimi or the Wig\-ner distributions, help to
reveal the structures of the corresponding classical phase space
\cite{TakahashiSaito85,BohigasTomsovicUllmo93,%
ChangShi85,ChangShi86,MirbachKorsch94,%
GorinKorschMirbach97}. In particular, for the case of regular classical
dynamics, the Husimi function of a (quasi)energy eigenstate is localized along
the corresponding quantizing torus; for chaotic motion, it is spread over the
entire chaotic layer. If the classical dynamics is mixed, quantum-mechanical
states can be classified as regular or chaotic according to their distribution
in phase space \cite{GorinKorschMirbach97}. Moreover, the phase-space
representation of the asymptotic state of a dissipative quantum map
exhibits the
structures of the corresponding classical attractor \cite{DittrichGraham87}.
However, these analogies find their limit in the Heisenberg uncertainty
principle. It does not allow for arbitrarily fine phase-space structures and
results in smearing on action scales below $h$.
%
\subsection{Classical attractor}

To describe the classical dissipative dynamics of the driven double well, we
add an Ohmic friction force $F_\gamma=-\gamma p$ to the conservative equations
(\ref{ddw:dotx}), (\ref{ddw:dotp}),
\begin{eqnarray}
\dot x &=& {1\over m}p , \label{ddw:dotx,diss} \\
\dot p &=& -\gamma p - {\partial V(x,t)\over\partial x}.
\label{ddw:dotp,diss}
\end{eqnarray}
Friction destroys the time-reversal symmetry (\ref{ddw:time_reversal})
of the conservative system. Accordingly, dissipation breaks the reflection
symmetry of the phase-space portrait with respect to the $x$-axis, found
at zero phase of the driving (cf.~Fig.~\ref{ddw:fig:poincare}).
\begin{figure}[t]
\centerline{
\psfig{width=6cm,figure=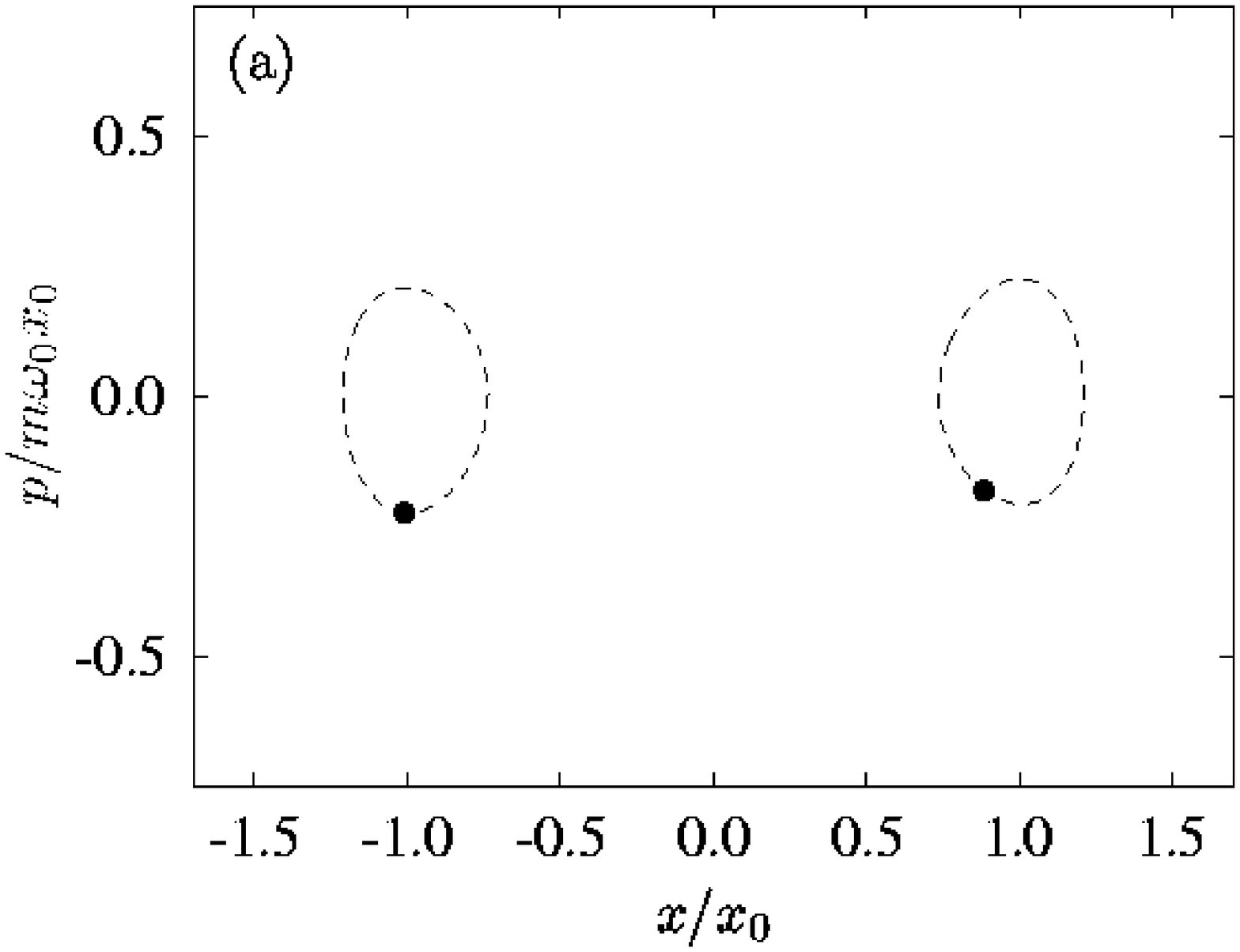}
\hfill
\psfig{width=6cm,figure=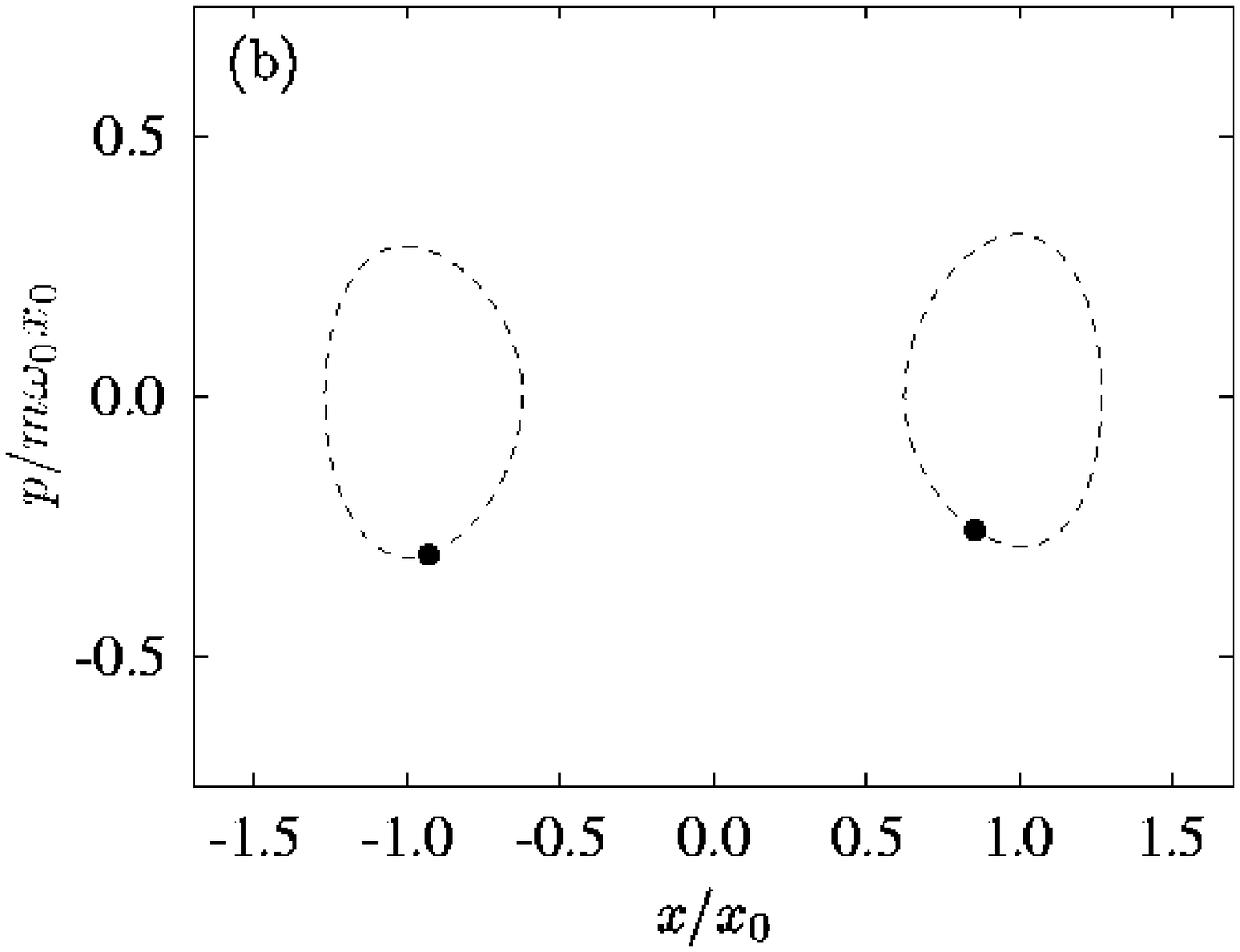}
}
\parbox[t]{6cm}{
\psfig{width=6cm,figure=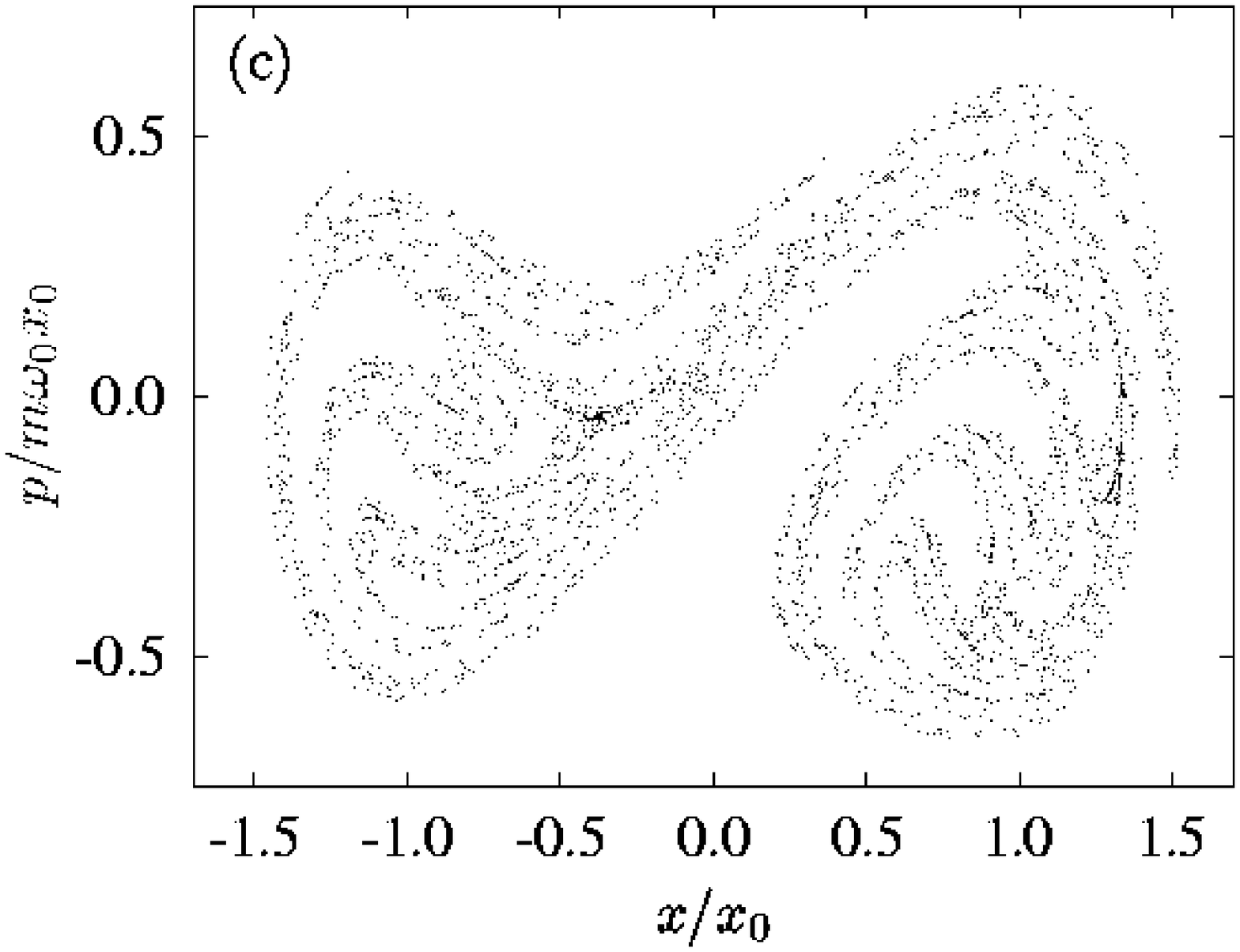}
}
\hfill\parbox[b]{5.2cm}{\caption{ \label{ddw:fig:poincare_diss}
Stroboscopic classical phase space portrait at $t=nP$, of the
dissipative harmonically driven quartic double well,
Eqs.\ (\protect\ref{ddw:dotx,diss}), (\ref{ddw:dotp,diss}),
for the driving amplitude $F=0.09$ and frequency $\Omega=0.9\,\omega_0$.
The friction strength is $\gamma=0.3\,\omega_0$ (a),
$0.2\,\omega_0$ (b), $0.03\,\omega_0$ (c).
In panels (a) and (b) the stroboscopic portrait is marked by a full dot and
the broken lines show the corresponding limit cycles.
}}
\end{figure}%
\begin{figure}[t]
\parbox[t]{6cm}{
\psfig{width=6cm,figure=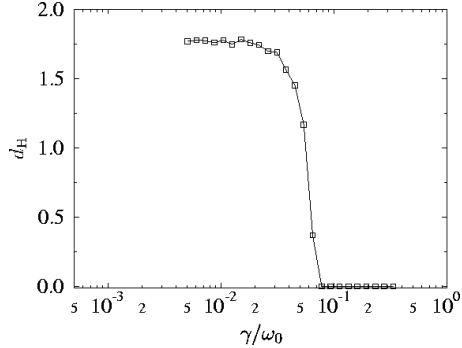}
}
\hfill\parbox[b]{5cm}{\caption{\label{ddw:fig:fracdim}
Hausdorff dimension of the classical attractor of the
dissipative harmonically driven quartic double well,
Eqs.~(\protect\ref{ddw:dotx,diss}), (\ref{ddw:dotp,diss}), for $F=0.09$,
$\Omega=0.9\,\omega_0$.
}\vspace{4ex}}
\end{figure}%

A constituent feature of dissipative flows is the net exponential
contraction of phase-space volume. Therefore, the dynamics is
asymptotically confined to an attractor, a measure-zero manifold in phase space
to which all trajectories starting from within the surrounding basin of
attraction converge. For periodically driven dissipative systems,
the attractor is in general also time-dependent with the period of the driving
and is adequately described by a stroboscopic map
\cite{MoonLi85a,MoonLi85b,Szemplinska92}.

Depending on the parameters that control the dissipative flow,
an attractor can consist of fixed points, limit cycles, or manifolds of fractal
dimension, less than that of phase space. An adequate concept to characterize
the geometry of an attractor is the Hausdorff dimension $d_{\rm H}$
defined, for example, in Ref.\ \cite{Ott93}.
It typically increases with decreasing
contraction rate, so that strange attractors are expected to occur in the
regime of weak dissipation of a system that in absence of friction, is chaotic.

The Hausdorff dimension of the classical attractor for the driven double well
with dissipation, Eqs.~(\ref{ddw:dotx,diss}), (\ref{ddw:dotp,diss}),
at the parameter values $F=0.09$ and
$\Omega=0.9\,\omega_0$, is shown in Fig.\ \ref{ddw:fig:fracdim} for various
values of the friction $\gamma$. Although the attractor itself is periodically
time dependent with the period of the driving, its Hausdorff dimension $d_{\rm
H}$ remains nearly constant \cite{MoonLi85a}. Near $\gamma\approx
0.06\,\omega_0$, with decreasing $\gamma$, the classical dynamics undergoes a
transition from regular motion to chaos, manifest in a corresponding transition
from limit cycles (Fig.~\ref{ddw:fig:poincare_diss}a,b)
to a strange attractor (Fig.~\ref{ddw:fig:poincare_diss}c)
and a concomitant jump in $d_{\rm H}$. At the same values of $F$ and $\Omega$,
the regular islands near the potential minima of the corresponding undamped
dynamics have already completely dissolved in the chaotic sea.

\subsection{Quantum attractor}

\begin{figure}[p]
\centerline{
\psfig{width=6cm,figure=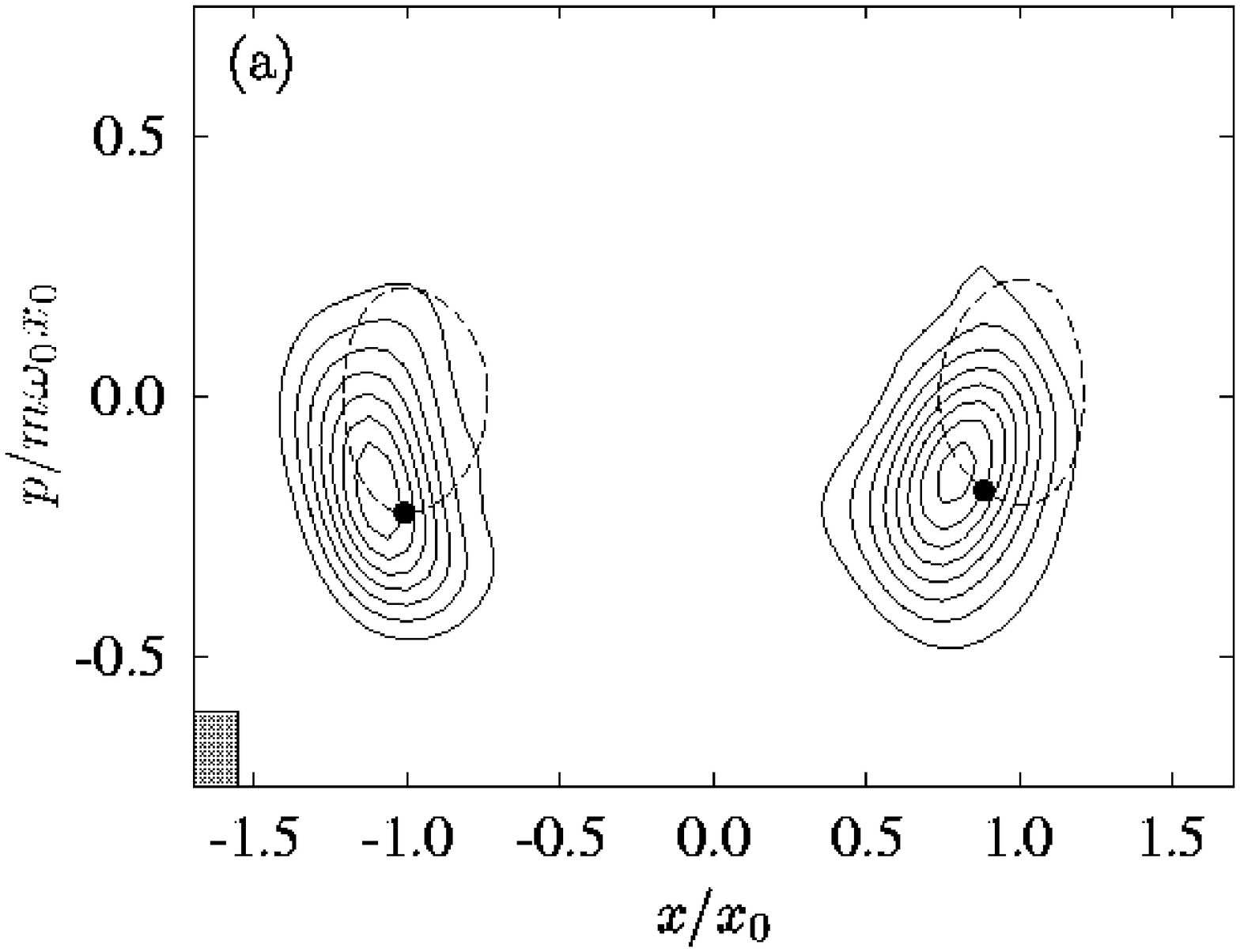}
\psfig{width=6cm,figure=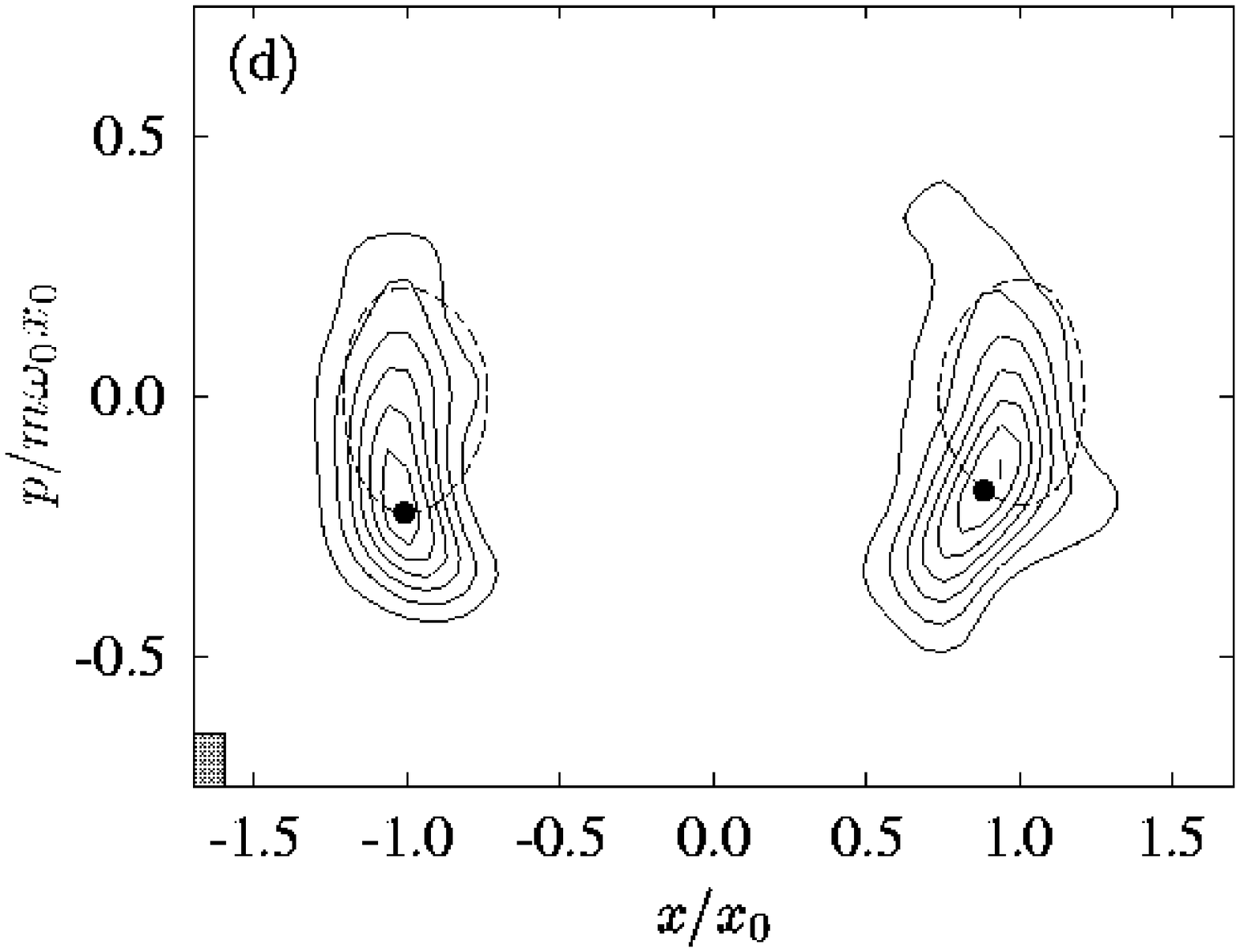}
}
\centerline{
\psfig{width=6cm,figure=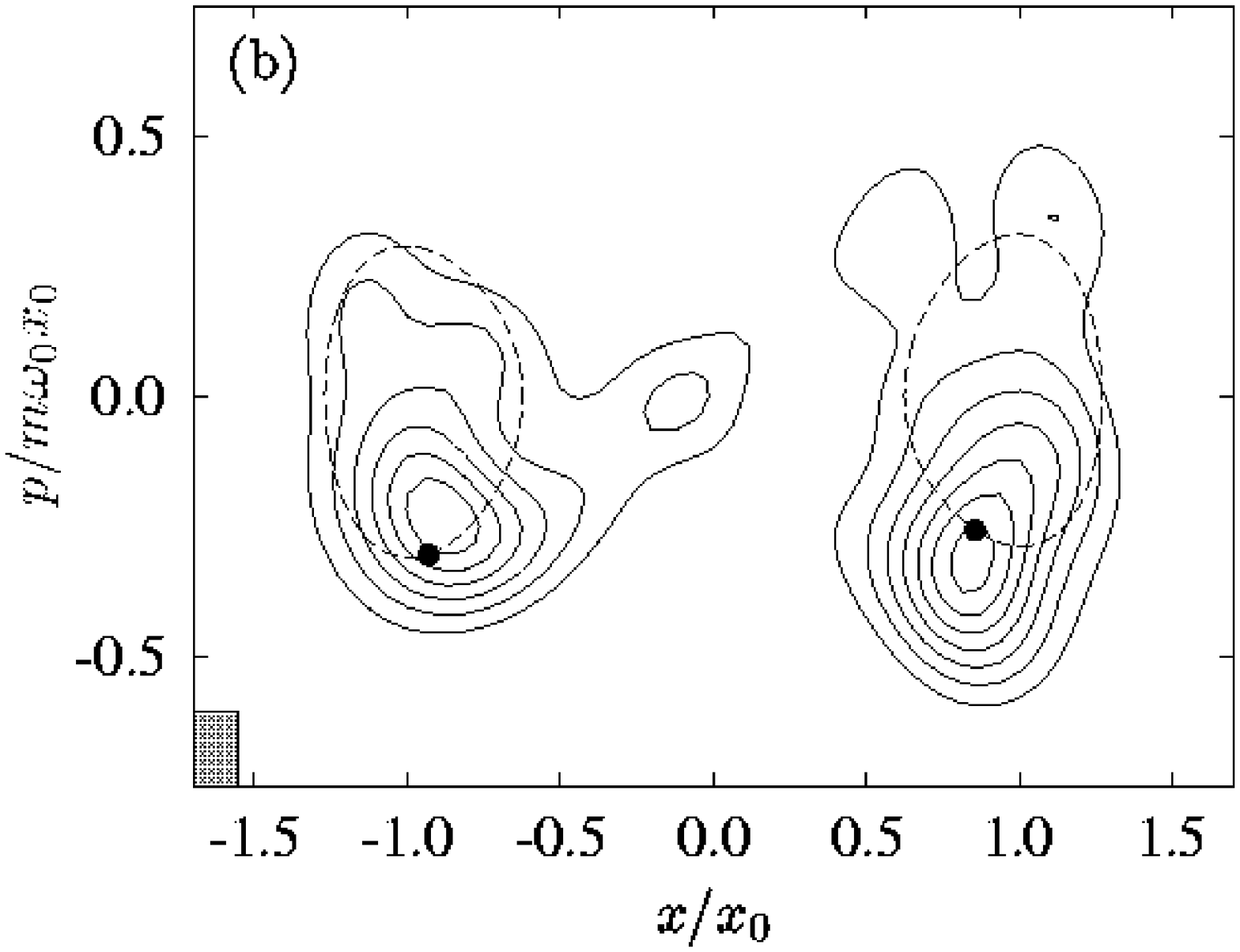}
\psfig{width=6cm,figure=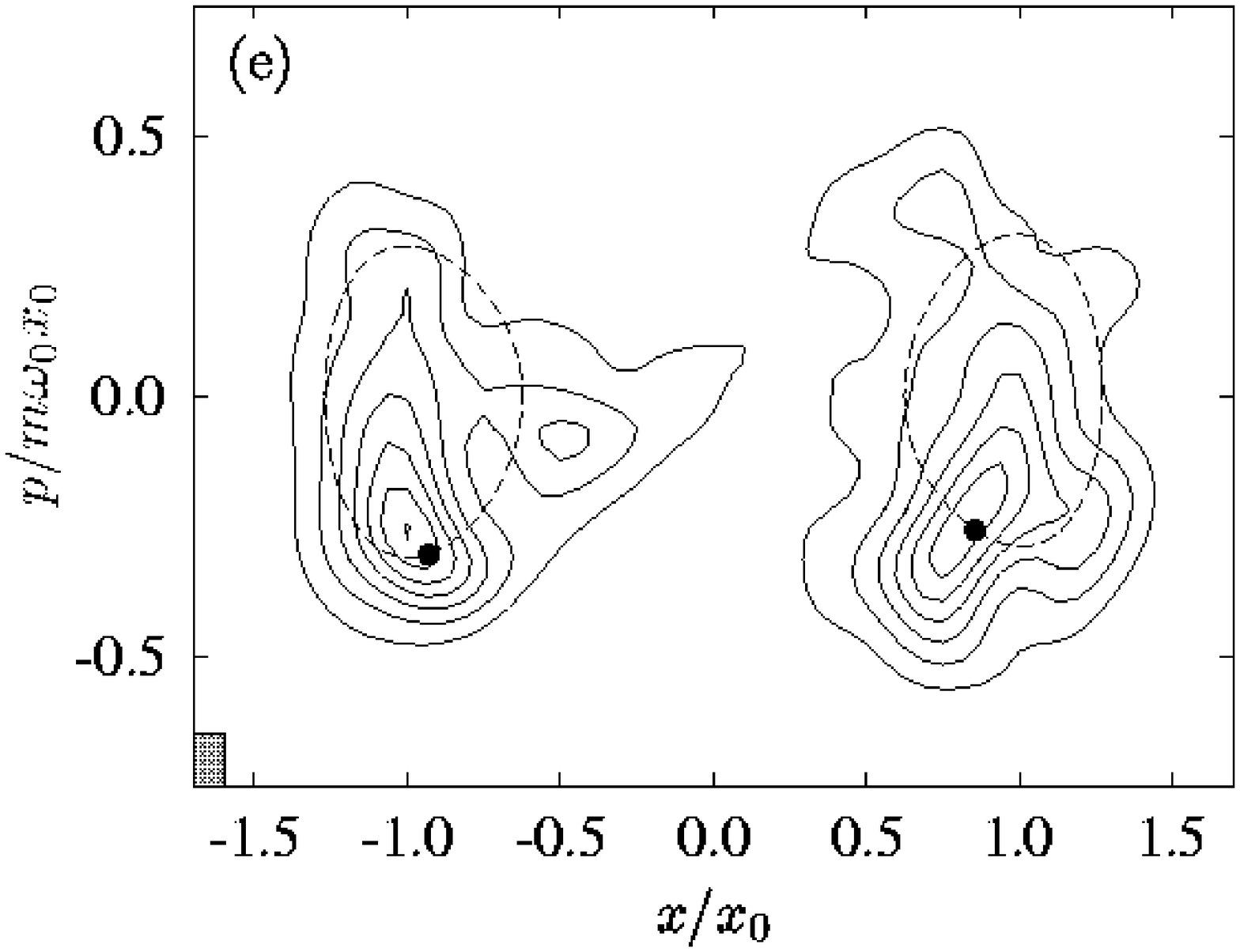}
}
\centerline{
\psfig{width=6cm,figure=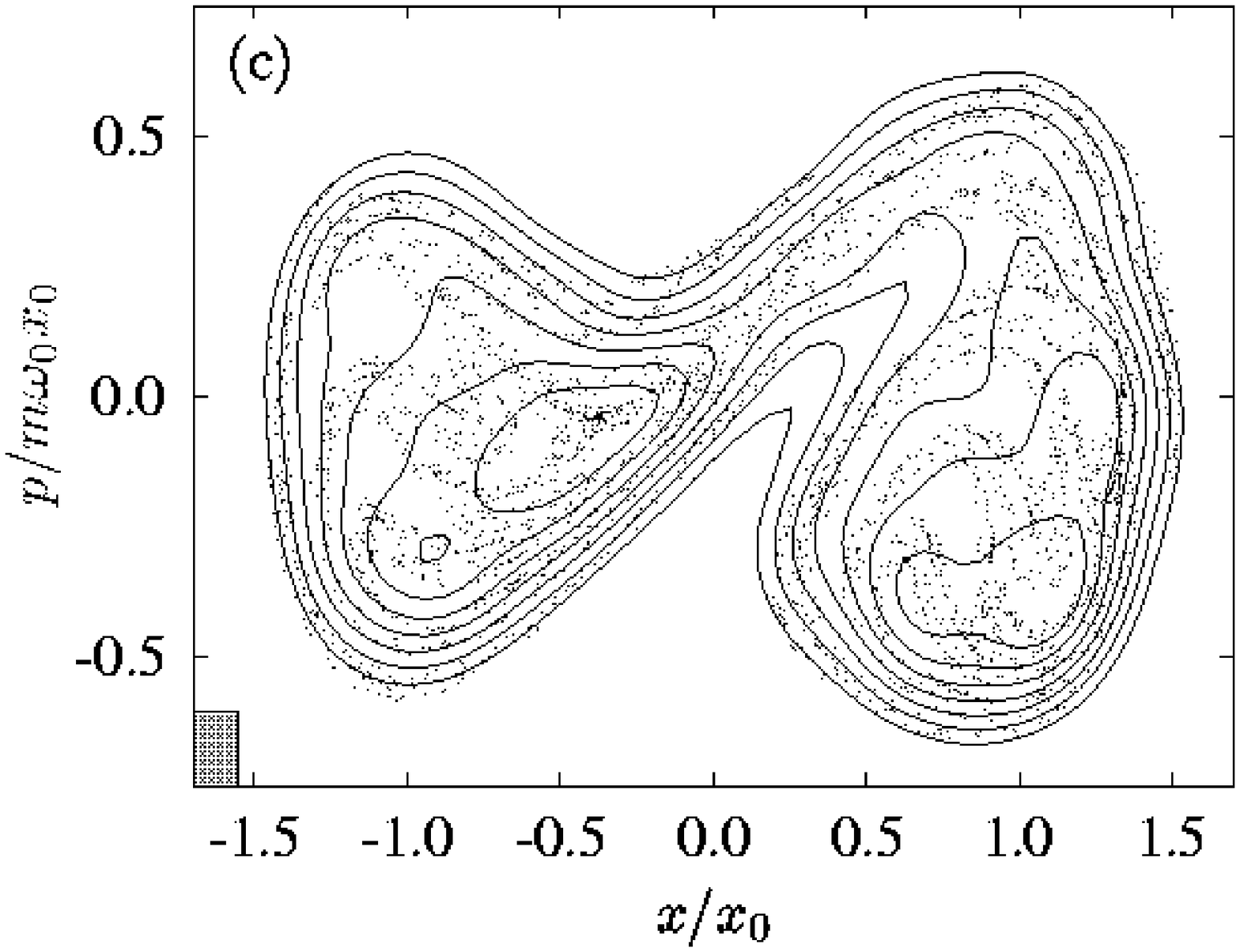}
\psfig{width=6cm,figure=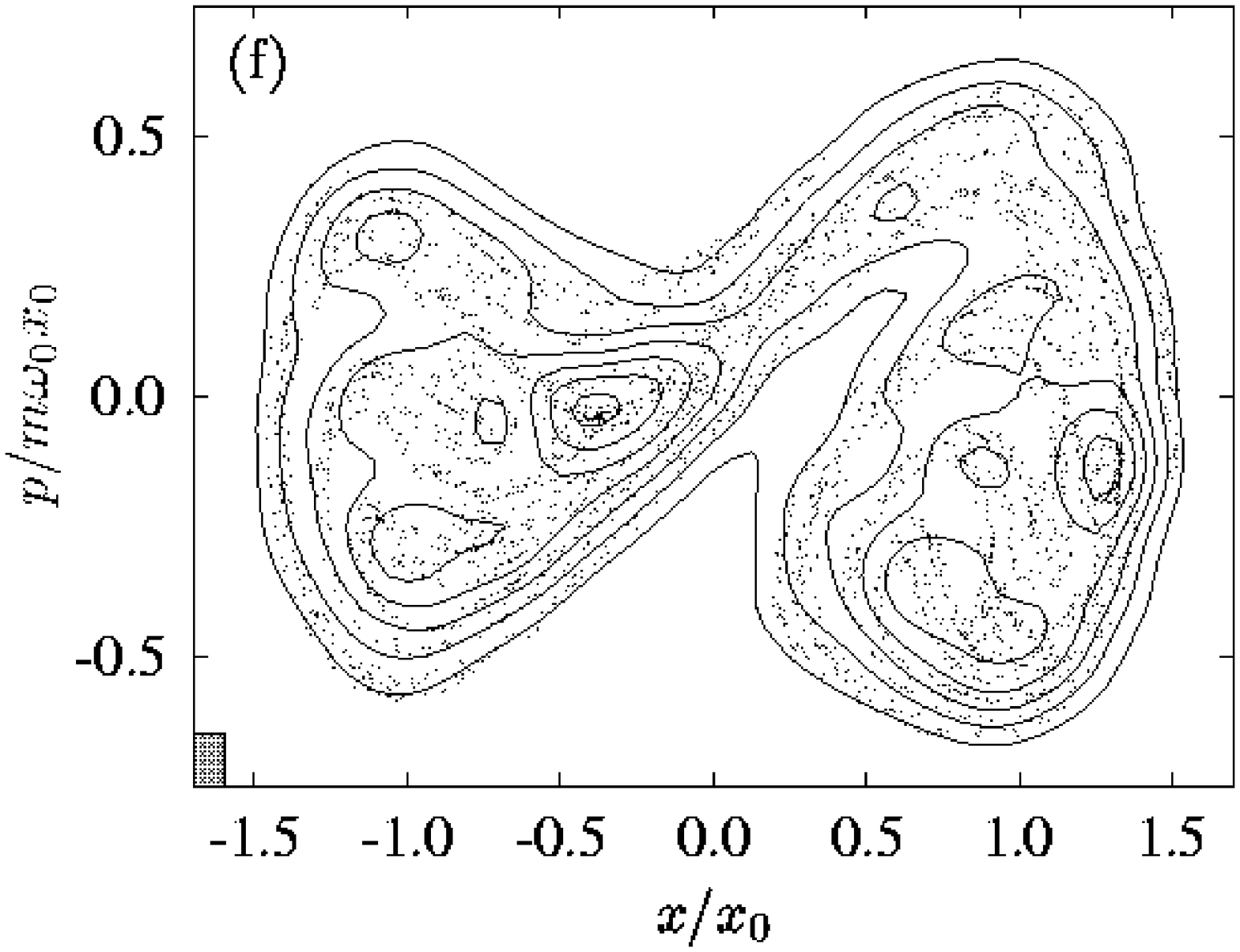}
}
\caption{ \label{ddw:fig:qattractor6+12}
Contour plot of the Husimi function of the quantum attractor
(full lines) at $t=nP$, $n\to\infty$,
superposed on the corresponding classical phase-space portrait,
Fig.\ \ref{ddw:fig:poincare_diss}.
The parameter values $F=0.09$, $\Omega=0.9\,\omega_0$,
$\gamma=0.3\,\omega_0$ (a,d), $0.2\,\omega_0$ (b,e),
$0.03\,\omega_0$ (c,f) are as in Fig.\ \ref{ddw:fig:poincare_diss}.
The effective action is $D=6$ (a--c) and $D=12$ (d--f).
The rectangle in the lower left corner has the size of the effective
quantum of action $\hbar_{\rm eff}=1/8D$.
}
\end{figure}%
\begin{figure}[t]
\parbox[t]{6cm}{
\psfig{width=6cm,figure=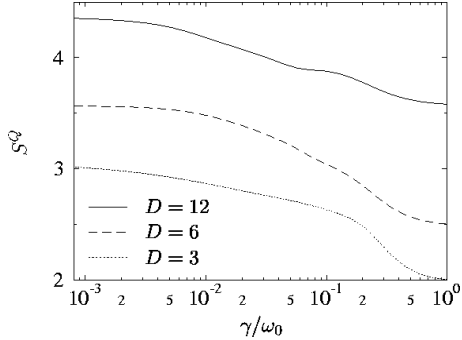}
}
\hfill\parbox[b]{5cm}{\caption{\label{ddw:fig:entropy}
Wehrl entropy of the asymp\-to\-tic state of the dissipative quantum map for
different values of the effective quantum of action $\hbar_{\rm eff}=1/8D$.
Other parameters as in Fig.\ \ref{ddw:fig:fracdim}.
}\vspace{2ex}}
\end{figure}
The ``quantum attractors'', i.e., the asymptotic states of the dissipative
quantum dynamics, for example in a Husimi representation, resemble the
corresponding classical attractors up to coarse graining
(Fig.~\ref{ddw:fig:qattractor6+12}). Correspondingly, the
qualitative transformation from limit cycles to a strange attractor is visible
in the asymptotic quantum distribution, but proceeds continuously. Although the
asymptotic state for $\gamma=0.2\,\omega_0$
(Fig.~\ref{ddw:fig:qattractor6+12}b,e), is still concentrated near the fixed
points of the classical stroboscopic map, it covers a broader phase-space
area that already anticipates the shape of the strange attractor.
Reducing the effective quantum of action $\hbar_{\rm eff}=1/8D$ allows for
a sharper resolution of the underlying classical structures in the Husimi
functions, as expected.

Like the attractors of the dissipative classical dynamics
(Fig.~\ref{ddw:fig:poincare_diss}), their quan\-tum-mechanical
counterparts have lost the reflection symmetry with respect to the $x$-axis,
in contrast to the Husimi representations of the Floquet states
in absence of dissipation (cf.\ Fig.\ \ref{ddw:fig:Qreg/chao}). This symmetry
breaking is caused by finite off-diagonal elements of the asymptotic density
matrix in Floquet representation, since diagonal representations share the
symmetries of the basis. Thus, off-diagonal matrix elements play a significant
r\^ole for the asymptotic state.
This demonstrates that a description within a full rotating-wave
approximation is insufficient, since it would result in an asymptotic
state diagonal in the Floquet representation
\cite{BlumelGrahamSirkoSmilanskyWaltherYamada89,%
BlumelBuchleitnerGrahamSirkoSmilanskyWalther91,GrahamHubner94}.

Because the quantum attractor, in contrast to the classical one, has support
all over phase space (or a region of finite measure), we cannot
characterize it by a Hausdorff dimension.
A more suitable measure for the extension of the quantum attractor
is a phase-space version of the Shannon entropy, the Wehrl entropy
\cite{GorinKorschMirbach97,Anderson93,Wehrl79}
\begin{equation}
S^Q = -\int \frac{{\rm d}x\,{\rm d}p}{2\pi\hbar}\, Q(x,p)\ln[Q(x,p)]
\end{equation}
of its Husimi representation
\begin{equation}
Q(x,p) = \langle x,p|\rho_{\rm S}|x,p\rangle ,
\end{equation}
where $|x,p\rangle$ denotes a coherent state centered at $(x,p)$ in
phase space.
Its exponential, $\exp(S^Q)$, gives approximately the number of coherent states
covered by the Husimi function. Thus, the occupied phase-space area is
$2\pi\hbar\exp(S^Q)$. The Wehrl entropy of the asymptotic state for our
numerical example is
depicted in Fig.\ \ref{ddw:fig:entropy} for various values of $\hbar_{\rm
eff}=1/8D$. It grows with decreasing friction $\gamma$, reflecting the
increasing dispersion of the Husimi functions. In the semiclassical regime,
i.e., for a sufficiently large value of the effective action $D$, we observe a
kink-like behavior of the entropy near $\gamma\approx0.06\,\omega_0$, where
the classical
attractor undergoes the transition mentioned above, from a set of isolated
fixed
points to a strange attractor.

Note that for
$\gamma\ga 0.1\,\omega_0$, the Markov approximation becomes inaccurate,
since $\gamma$ is then of the order of the mean level spacing and
the condition (\ref{mar:cond_frequency}) is violated for at least
part of the transitions between Floquet states.
Nevertheless, we obtain the qualitative behavior which we expected from
classical considerations.

\section{Conclusion}

For the generic situation of the dissipative quantum dynamics of
a particle in a driven double-well
potential, classical chaos plays a significant r\^ole for the coherent
dynamics.  Even for arbitrarily small driving amplitude, the separatrix is
replaced by a chaotic layer, while the motion near the bottom of the wells
remains regular. Nevertheless, the influence of states located in the
chaotic region alters the splittings of the regular doublets and thus
the tunnel rates, which is the essence of chaotic tunneling.
We have studied chaotic tunneling in the vicinity of crossings of chaotic
singlets with tunnel doublets under the influence of an environment.
As a simple intuitive model to compare against, we have constructed a
three-state system which in the case of vanishing dissipation, provides
a faithful description of an isolated singlet-doublet crossing.
Dissipation introduces new time scales to the system: one for the loss of
coherence and a second one for the relaxation to an asymptotic state.
Well outside the crossing, both time-scales are of the same order, reflecting
an effective two-state behavior. The center of the crossing is characterized
by a strong mixing of the chaotic state with one state of the tunnel doublet.
The high mean energy of the chaotic state introduces additional decay channels
to states outside the three-state system. Thus, decoherence becomes
far more effective and, accordingly, tunneling fades out much faster.

The study of the asymptotic state, the quantum attractor, demonstrates clearly
that a three-state model of the singlet-doublet crossing is insufficient once
dissipation is effective. This is so because the coupling to the
heat bath enables processes of decay and thermal activation that connect the
states in the crossing with other, ``external'' states of the central
system. In the presence of driving, the asymptotic state is no longer
literally a state of equilibrium. Rather, incoherent processes create a steady
flow of probability involving states within as well as outside the
crossing. As a result, the composition of the asymptotic state, expressed for
example by its coherence ${\rm tr}\,\rho_\infty^2$, is markedly
different at the center of the crossing as compared to the asymptotic state
far away from the crossing, even if that is barely visible in the corresponding
classical phase-space structure.

With increasing driving amplitude, in absence of dissipation, even the dynamics
near the bottom of the wells becomes fully chaotic.
This has striking consequences for the corresponding dissipative classical
dynamics: For sufficiently weak dissipation, it remains chaotic, but for strong
friction it becomes regular. Accordingly, the geometry of the classical
attractor is fractal or regular, respectively.
We have observed the signatures of this qualitative difference in the
asymptotic state of the corresponding quantum dynamics.
However, in contrast to the sudden change of the classical behavior, the
quantum attractor undergoes a smooth transition:
The structure of the strange attractor is already felt by the Husimi
function for parameter values where the classical attractor consists only
of two isolated fixed points.
For the observation of these semiclassical structures, off-diagonal matrix
elements of the asymptotic state in Floquet basis proved crucial.
This clearly indicates that a full rotating-wave approximation must fail.

Many more phenomena at the overlap of chaos, tunneling, and dissipation await
being unraveled. They include four-state crossings formed when two doublets
intersect, chaotic Bloch tunneling along extended potentials with a large
number of unit cells instead of just two, and the influence of decoherence on a
multi-step mechanism of chaotic tunneling.
These phenomena are typically observed in the far semiclassical regime,
which requires to take very many levels into account.
A semiclassical description of the dissipative quantum system may circumvent
this problem.

%

\end{document}